\documentclass[aps,prb,twocolumn,groupedaddress]{revtex4-1}

\usepackage{graphicx}

\begin{document}



\title{Revisiting the electronic phase diagram of YBa$_2$Cu$_3$O$_y$ via temperature derivative of in-plane resistivity}



\author{Yue Wang}
\email[]{yue.wang@pku.edu.cn}
\affiliation{State Key Laboratory for Mesoscopic Physics, School of Physics, Peking University, Beijing 100871, China}

\author{Jun Zhang}
\affiliation{State Key Laboratory for Mesoscopic Physics, School of Physics, Peking University, Beijing 100871, China}


\date{\today}

\begin{abstract}
  In-plane resistivity $\rho_{ab}$ has been a helpful route to delineate the electronic phase diagram of high-$T_c$ cuprate superconductors. In this paper we have re-examined the temperature $T$ versus doping $p$ phase diagram of YBa$_2$Cu$_3$O$_{y}$ (YBCO) to address several issues therein by using the temperature derivative of in-plane resistivity, d$\rho_{ab}$/d$T$. It is found that the systematic evolution of d$\rho_{ab}$/d$T$ with $T$ and $p$ shows consistent features among YBCO polycrystals, thin films, and single crystals, allowing one to identify properties intrinsic to this material. For superconducting samples with $p$ less than about 0.15, a temperature $T_f$ has been defined to mark the onset of an upturn in d$\rho_{ab}$/d$T$ at $T>T_c$, which, in light of prior studies on another cuprate family La$_{2-x}$Sr$_{x}$CuO$_{4}$, is attributed to the onset of superconducting fluctuations in the normal state. The $T_f$, in contrast to the well-documented pseudogap temperature $T^\ast$, exhibits a doping dependence similar to $T_c$, and the interval between $T_f$ and $T_c$ is about $5-30$ K across the underdoped regime, showing agreement with a variety of other types of measurements to probe a restricted temperature range of superconducting fluctuations in the pseudogap phase of YBCO. Above $T_f$, the d$\rho_{ab}$/d$T$ increases linearly as temperature increases up to a value denoted as $T_2$, which is about half the $T^\ast$ and falls roughly in parallel with $T^\ast$ as doping rises up to about 0.13, indicating a prominent $T^2$-dependent $\rho_{ab}$ in this $T-p$ region. The detailed doping evolution of d$\rho_{ab}$/d$T$ helps further reveal that, at $T_f<T<T_2$, the $\rho_{ab}$ also involves an insulating-like component characterized approximately by a $\ln(1/T)$ dependence for $p<0.08$, or a $T$-linear component for $p>0.10-0.11$, thereby leaving a narrow window of doping for $\rho_{ab}$ to show a pure $T^2$ dependence. As $T$ increases further, the d$\rho_{ab}$/d$T$ reaches a local maximum at a temperature $T_R$, signifying the known curvature change (inflection point) in $\rho_{ab}(T)$. By also plotting the derivatives, it is illustrated that, in the vicinity of the inflection-point temperature $T_R$, the in-plane Hall coefficient $R_\mathrm{H}(T)$ and thermopower $S_{ab}(T)$ of YBCO display curvature changes as well, suggesting a correlation of the three transport properties. In addition, it is found that, the d$S_{ab}$/d$T$ also shows the onset of an upturn at a temperature coinciding with the $T_f$, which, backing the identification of $T_f$ with the onset of superconducting fluctuations, demonstrates further the virtue of using the temperature derivative to unveil information helpful for the study of high-$T_c$ cuprates.
\end{abstract}





\maketitle


\section{\label{sec:introduction}Introduction}

High transition-temperature cuprate superconductors exhibit a rich phase diagram in the plane of temperature $T$ versus carrier doping $p$, understanding of which is arguably a necessary step towards identifying the mechanism of superconductivity in these materials. \cite{Norman2011,Keimer2015} One prominent anomaly shown in the phase diagram is the presence of a pseudogap opening in the normal state at a characteristic temperature $T^\ast$ well above the superconducting transition temperature $T_c$. \cite{Timusk1999,Norman2005} As $p$ diminishes from the overdoped side, although the $T_c$ begins falling when $p$ reducing below the optimal-doping point, the $T^\ast$ continues to grow to higher temperatures, signifying a wider temperature range of the pseudogap phase towards underdoping. \cite{Timusk1999} Currently, the nature of the pseudogap state has remained a subject of debate. \cite{Norman2011,Keimer2015} Some experiments reported the broken time-reversal symmetry \cite{Fauque2006,Mook2008,Li2008} or broken rotational symmetry \cite{Daou2010} at temperatures below $T^\ast$, suggesting the bonding of pseudogap with a line of phase transitions. \cite{Shekhter2013} While in other experiments, similar momentum and doping dependences of the pseudogap and superconducting gap and a smooth evolution of the pseudogap into superconducting gap as $T$ decreases across $T_c$ have been observed, \cite{Renner1998,Norman1998,Meng2009} indicative of a precursor superconducting state of the pseudogap phase. \cite{Emery1995} To examine this latter possibility, one needs to experimentally probe the fluctuating superconductivity above $T_c$ and compare the temperature at which it vanishes with the $T^\ast$ across the phase diagram. In this respect, we note that conflicting results on the temperature range of superconducting fluctuations above $T_c$ have been obtained by using different experimental techniques in certain cuprate families. In La$_{2-x}$Sr$_x$CuO$_{4}$ (LSCO), for instance, vortex-like Nernst signals and diamagnetism were observed to persist well above $T_c$ in Nernst effect and torque magnetometry measurements, \cite{Wang2006,Li2010} suggesting the survival of superconducting correlations far into the pseudogap phase, whereas in recent terahertz spectroscopy measurements signatures of superconducting fluctuations were found to be present at most 16 K above $T_c$, i.e., in a rather narrower temperature range and well below $T^\ast$. \cite{Bilbro2011}

In-plane resistivity of high-$T_c$ cuprates has long been scrutinized to help elucidate their underlying electronic phase diagram. \cite{Hussey2008} In earlier studies, it has been found that the pseudogap temperature $T^\ast$ determined by probes such as nuclear magnetic resonance coincides with the temperature at which the resistivity starts to deviate from its approximate linear-$T$ behavior in the high-temperature regime. \cite{Timusk1999} Such findings suggested the influence of the development of pseudogap on charge transport properties and made it a convenient method to define the $T^\ast$ by investigating temperature and doping evolution of the in-plane resistivity $\rho_{ab}$. Recently, it was demonstrated that the onset of superconducting fluctuations above $T_c$ could also be identified from the resistivity data. \cite{Rourke2011} In overdoped LSCO, it has been shown that the onset temperature for superconducting fluctuations, $T_{f}$, marked by the temperature at which the magnetoresistance begins to deviate from its well-established behavior in the normal state, correlates precisely with the temperature where a clear upturn in d$\rho_{ab}$/d$T$, i.e., the temperature derivative of in-plane resistivity, appears. \cite{Rourke2011} This illustrates the usefulness of the derivative of planar resistivity in uncovering the temperature range of superconducting fluctuations above $T_c$. In fact, it has been noted that, by taking the first or second derivative of in-plane resistivity, other important features inherent in the resistivity curve, such as the temperature dependence of the resistivity, may become more transparent as well. \cite{Ando2004,Hussey2011} In particular, by using this method,\cite{Cooper2009,Hussey2011,Hussey2013} it has been recently shown that, for overdoped high-$T_c$ cuprates the normal-state resistivity at low temperatures could actually be decomposed into two $T$-dependent components, a $T$-linear plus a $T^2$ component. \cite{Hussey2008,Cooper2009} In overdoped LSCO, the magnitudes of these two $T$-dependent terms were found to show different variations with doping and both displayed a change as the doping decreased across $p\simeq0.19$, suggesting the presence of a critical point in the doping evolution of resistivity. \cite{Cooper2009} As doping further decreased into the underdoped regime, whether the resistivity could still be delineated as a sum of such two $T$-dependent components and how their strengths continued to evolve with doping are therefore worth exploring. Moreover, in recent years some new electronic ordering phenomena have been discovered within the pseudogap phase of high-$T_c$ cuprates. \cite{Norman2011,Keimer2015} For instance, recent x-ray diffraction experiments have detected incommensurate charge order setting in at elevated temperatures below $T^\ast$ in several cuprate families. \cite{Keimer2015,Comin2016} What is the potential impact of such electronic orders on the charge transport? Could it be identified in resistivity or other transport properties of cuprates? A revisit of the in-plane resistivity may also be helpful to shed light on these intriguing issues.

Motivated by such recent progresses and discoveries, in this paper we have re-examined the temperature and doping evolution of in-plane resistivity in the archetypal cuprate YBa$_2$Cu$_3$O$_{y}$ (YBCO), with the purpose of identifying the temperature range of superconducting fluctuations and also investigating in detail the temperature dependence of $\rho_{ab}$ in the pseudogap phase. We start with the fabrication of YBCO bulk samples and the measuring of resistivity as a function of temperature at different dopings, as described in Sec. \ref{sec:experiment}. Then in Sec. \ref{subsec:Res-Deriv} we compare the resistivity in these polycrystals with that typically reported in YBCO thin films and single crystals, by focusing on the temperature and doping variation of the first derivative of resistivity. The essential features of $\rho_{ab}$ in YBCO become transparent through this comparison and several characteristic temperatures are defined to place them in the $T-p$ phase diagram. Sec. \ref{subsec:SC-fluc} is devoted to discussing one of the characteristic temperatures, namely the onset temperature for superconducting fluctuations $T_f$. It is shown that the $T_f$ yielded from resistivity is well below $T^\ast$ and follows a doping dependence like the $T_c$, and the width of superconducting fluctuations is relatively small across the phase diagram, less than about 30 K even at very low doping concentrations. A comparison of the $T_f$ from resistivity to that from a set of other experimental techniques is also shown, which seems to further back the notion of a narrow temperature range of superconducting fluctuations above $T_c$ in YBCO. In Sec. \ref{subsec:T-square-Res} we examine in detail another feature of $\rho_{ab}$ as manifested by the d$\rho_{ab}$/d$T$, namely the presence of $T^2$-dependence within the pseudogap phase. It is shown that the temperature $T_2$, below which a predominant $T^2$ dependence emerges in $\rho_{ab}$, is about half the $T^\ast$ and decreases roughly in parallel with $T^\ast$ as doping increases in the underdoped regime. Furthermore, it is shown that, prior to the display of a pure $T^2$ dependence at $0.08\lesssim p\lesssim0.10$, the $\rho_{ab}$ at lower dopings exhibits in $T_f<T<T_2$ the contribution of an insulating-like component which diminishes gradually as $p$ approaches 0.08. For $p>0.10$, while in some single crystals the $\rho_{ab}$ retains a pure $T^2$ dependence with $p$ increasing up to about 0.11, in more other samples a $T$-linear component begins to appear in $\rho_{ab}$ coexisting with the $T^2$-dependent one. Implications of this detailed doping evolution of $\rho_{ab}$ on low-$T$ transport properties of YBCO have been discussed. In Sec. \ref{subsec:Res-Hall-Seebeck} the in-plane Hall coefficient $R_\mathrm{H}(T)$ and thermopower $S_{ab}(T)$ of YBCO as typically reported are also examined by plotting the derivatives. It is shown that, over certain doping intervals, the temperature $T_2$ matches the temperature where the $S_{ab}(T)$ attains a maximum, and moreover, above $T_2$, in the vicinity of the temperature $T_R$ where the $\rho_{ab}(T)$ exhibits curvature change (inflection point), the $R_\mathrm{H}(T)$ and $S_{ab}(T)$ also exhibit the curvature change, suggesting a correlation among the three transport properties in the pseudogap phase. To gain insight into what may lie behind this observed phenomenon, a comparison of the characteristic temperatures shown in three transport properties to those identified recently in other kinds of experiments, such as the onset temperature of charge-density-wave order in x-ray diffraction, has been made and briefly discussed. We also show that the appearance of superconducting fluctuations in the normal state could as well be spotted in the temperature derivative of $R_\mathrm{H}(T)$ and particularly $S_{ab}(T)$, similar to the case with $\rho_{ab}(T)$. These further demonstrate the merit of plotting the temperature derivative to unveil useful information for the study of high-$T_c$ cuprates. In the end, we summarize the main findings in Sec. \ref{sec:summary}.

\section{\label{sec:experiment}Experiment}

We have prepared YBCO bulk samples via the conventional method of solid-state reaction. \cite{Pathak2005} Appropriate amount of Y$_{2}$O$_{3}$, BaCO$_{3}$, and CuO powders following the stoichiometry of YBCO were thoroughly mixed, calcined in air at $860-920$~$^{\circ}$C for 72 h with intermediate grindings, pelletized, and then sintered at 970~$^{\circ}$C for 48 h. The phase purity of obtained YBCO samples has been checked by powder x-ray diffraction measurements. To tune the oxygen contents, the as-synthesized samples of dimensions $\sim4\times0.6\times0.2$~mm$^3$ have been annealed under various conditions, similar to those described by others. \cite{Gao2006} To achieve near optimum-doping point, the sample has been annealed in flowing oxygen at 550~$^{\circ}$C for 72 h and cooled slowly at a rate of 100~$^{\circ}$C/h down to room temperature. To obtain underdoped samples, on the other hand, the annealing at temperatures of $500-900$~$^{\circ}$C for $24-48$ h in flowing oxygen or argon has been followed by quenching the sample quickly into liquid nitrogen.

The electrical resistivity measurement has been performed in a physical property measurement system (PPMS, Quantum Design) by using the standard four-probe method. Figure \ref{fig-RT} summarizes the temperature dependence of resistivity for samples with hole doping concentration $p$ spanning from 0.084 to 0.152. The doping $p$ has been evaluated from the $T_c$ of the samples according to the established relation between the two for YBCO. \cite{Liang2006} Here, the $T_c$ has been defined as the midpoint of the superconducting transition.

Owing to the polycrystalline nature of our sample, in resistivity measurement there should be grains in the sample not aligned in the $ab$-plane in direction of the applied electrical current, resulting in the measured resistivity involving both in-plane and out-of-plane ($c$-axis) contributions. Whether this would affect us to extract the intrinsic property of the in-plane resistive transport in YBCO therefore needs to be checked, although it has been well documented that the magnitude of the $c$-axis resistivity $\rho_c$ is much (hundred times) larger than the in-plane resistivity $\rho_{ab}$ in YBCO owing to its layered structure and highly two-dimensional electronic state, \cite{Takenaka1994} implying that the resistivity measured in polycrystalline YBCO may still mainly reflect the in-plane transport property of the sample by assuming a simple parallel conduction of both in- and out-of-plane directions. To examine this, and more importantly to seek universal features shown in the in-plane resistivity of YBCO, we have revisited typical results of $\rho_{ab}$ obtained previously on YBCO epitaxial thin films and single crystals, and compared them with the present data on polycrystalline samples, as described in detail in the following.

\section{\label{sec:results}Results and discussion}

\subsection{\label{subsec:Res-Deriv}Doping evolution of the in-plane resistivity and its temperature derivative}

\begin{figure}
  \includegraphics[scale=0.3]{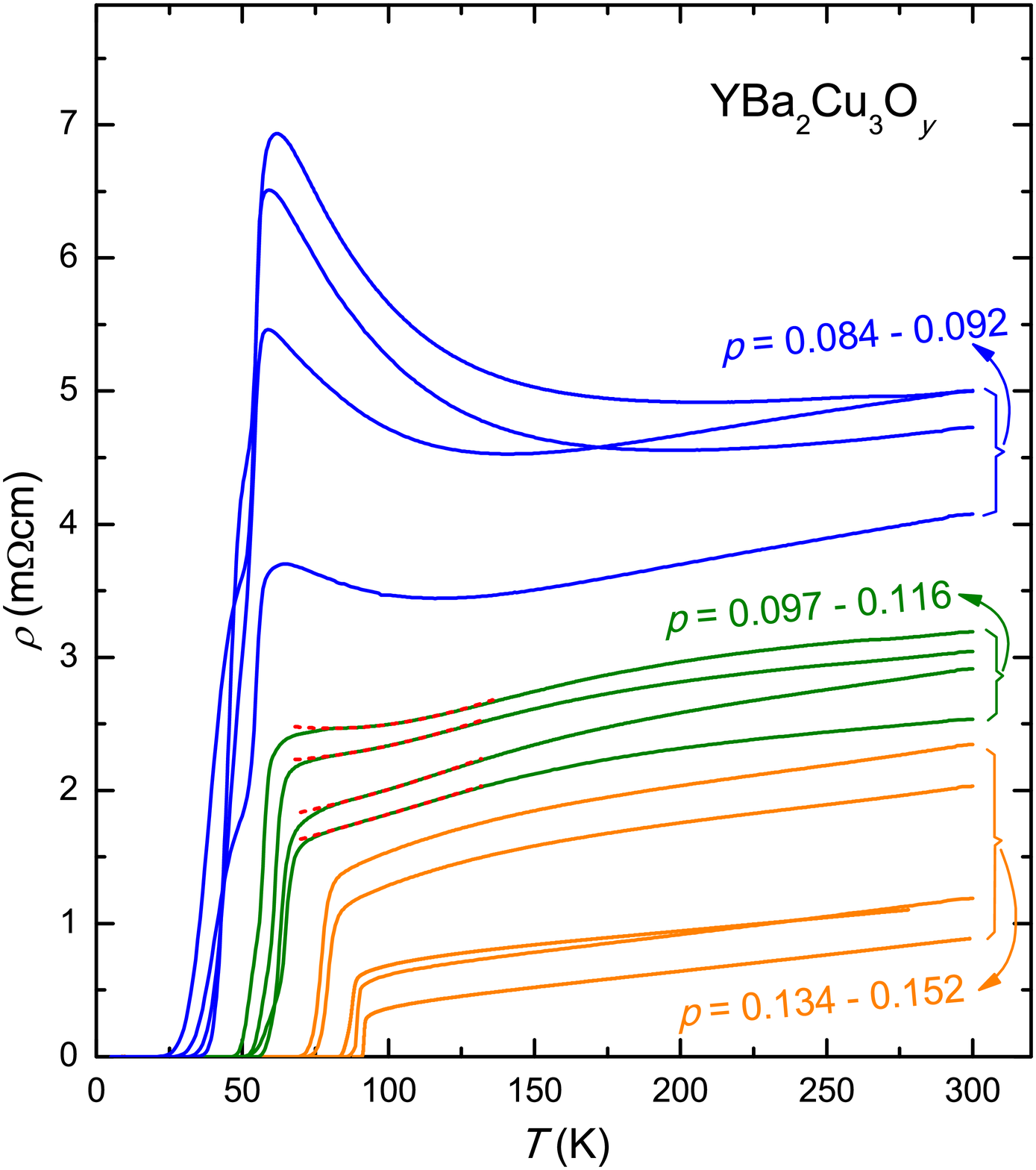}
  \caption{\label{fig-RT}(Color online) Temperature dependence of the resistivity $\rho$ for YBa$_2$Cu$_3$O$_y$ polycrystals with hole doping concentration $p$ varied from 0.084 to 0.152. At intermediate dopings, the $\rho(T)$ curves (green) exhibit an upward curvature at low temperatures above $T_c$. In this temperature regime, the derivative of resistivity [Figs. \ref{fig-Deriv}(e) and \ref{fig-Deriv}(f)] reveals presence of a $T^2$-dependent component in the resistivity, as illustrated by red dashed lines which are fits to data according to Eqs. (\ref{eq-doping-regime-1}) and (\ref{eq-doping-regime-3}). See the text for details.}
\end{figure}

\begin{figure*}
  \includegraphics[scale=0.61]{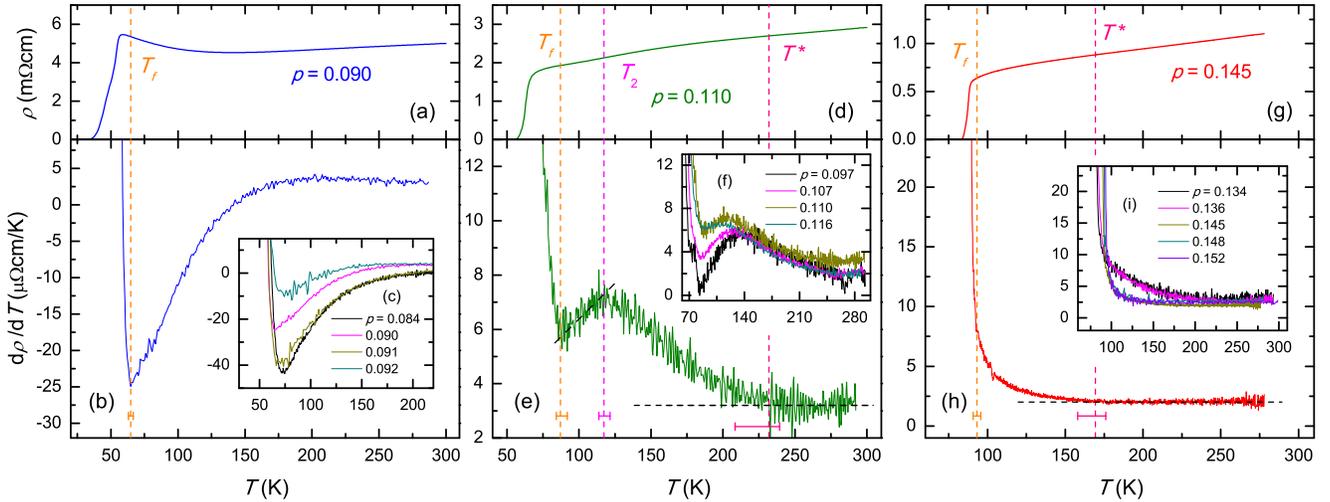}
  \caption{\label{fig-Deriv}(Color online) Doing evolution of temperature derivative d$\rho$/d$T$ of the resistivity shown in Fig. \ref{fig-RT} for YBa$_2$Cu$_3$O$_y$ polycrystalline samples. Left: the resistivity for doping $p=0.090$ (a) and its temperature derivative (b). The vertical dashed line marks the minimum in d$\rho$/d$T$, defined as the onset temperature for fluctuating superconductivity $T_f$. (c) plots the d$\rho$/d$T$ for the four low-doping samples showing qualitatively similar temperature dependences. Middle: the resistivity of $p=0.110$ (d) and its temperature derivative (e), as a representative of the d$\rho$/d$T$ behavior of the four intermediate-doping samples (f). At high $T$, the d$\rho$/d$T$ is nearly constant, as illustrated by a horizontal dashed line, reflecting a $T$-linear dependence of the resistivity. As $T$ drops, the d$\rho$/d$T$ starts to grow and deviate from this $T$-constant behavior at a temperature defined as the pseudogap temperature $T^\ast$. As $T$ is further reduced, an additional temperature scale labelled as $T_2$ is defined, below which the d$\rho$/d$T$ decreases linearly with decreasing $T$ (as indicated by a diagonal dashed line). It suggests the resistivity contain a $T^2$-dependent term for temperatures between $T_2$ and $T_f$ [defined as in (b)]. Right: the resistivity of $p=0.145$ (g) and its temperature derivative (h), typical of the d$\rho$/d$T$ behavior of the five relatively high-doping samples (i). The $T^\ast$ is defined as in (e). Below $T^\ast$, the d$\rho$/d$T$ increases all the way with decreasing $T$, with no intervening decrease-in-$T$ behavior as shown in (e). Nevertheless, a close look at the d$\rho$/d$T$ curve at low $T$ reveals a kink feature in its increase behavior with decreasing $T$. The $T_f$ is located at this kink temperature below which the d$\rho$/d$T$ grows much steeper with reducing temperature. The horizontal error bars in (b), (e), and (h) are to indicate the uncertainty in determining the temperatures $T_f$, $T_2$, and $T^\ast$.}
\end{figure*}

In Fig. \ref{fig-RT}, all the $\rho(T)$ curves of YBCO polycrystalline samples have been shown as three groups with variation of the doping. This is made based on the temperature derivative of the resistivity d$\rho$/d$T$ as displayed in Fig. \ref{fig-Deriv}, with the $\rho(T)$ curves in each group showing qualitatively similar d$\rho$/d$T$ behaviors. When doping is between 0.084 and 0.092, Fig. \ref{fig-RT} shows that the $\rho(T)$ exhibits a low-$T$ upturn before the superconducting transition sets in, suggesting a localization-like charge transport at low temperatures similar to the observation in lightly-doped high-$T_c$ cuprates. \cite{Ando2001} Accordingly, in this temperature regime the d$\rho$/d$T$ shows negative values and decreases with reducing $T$, as shown in Figs. \ref{fig-Deriv}(b) and \ref{fig-Deriv}(c), until the temperature reduces to a value denoted as $T_f$, below which the d$\rho$/d$T$ no longer decreases and, on the contrary, increases sharply, yielding a well-defined minimum in d$\rho$/d$T$. In view of previous study, \cite{Rourke2011} the $T_f$ is identified as the onset temperature for superconducting fluctuations, as will be discussed later. As doping increases to be between 0.097 and 0.116, Fig. \ref{fig-RT} shows that the $\rho(T)$ keeps metallic at all $T$ in the normal state, with d$\rho$/d$T>0$ at $T>T_c$ as demonstrated in Figs. \ref{fig-Deriv}(e) and \ref{fig-Deriv}(f). To capture main features of the d$\rho$/d$T$ curves for these intermediate-doping samples, as shown in Fig. \ref{fig-Deriv}(e), besides the temperature $T_f$ as defined in Fig. \ref{fig-Deriv}(b), two other characteristic temperatures labelled as $T^\ast$ and $T_2$ have also been defined. Below $T^\ast$, the d$\rho$/d$T$ deviates from its nearly $T$-constant behavior at high temperatures, suggesting a deviation of $\rho(T)$ from a $T$-linear dependence. Hence, following the standard definition, \cite{Timusk1999} the $T^\ast$ denoted is the pseudogap temperature identified from resistivity. The $T_2$ is a lower temperature scale, marking the point below which the d$\rho$/d$T$ is shown to decrease approximately linearly with decreasing temperature. This indicates that the $\rho(T)$ contain a $T^2$-dependent component between $T_2$ and $T_f$. As shown in Fig. \ref{fig-Deriv}(f), the slope of the decrease in d$\rho$/d$T$ in this temperature range reduces as the doping is increased, suggesting a decline of the strength of this $T^2$-dependent term towards higher doping levels. When doping rises further to be between 0.134 and 0.152, Figs. \ref{fig-Deriv}(h) and \ref{fig-Deriv}(i) show that the d$\rho$/d$T$ exhibits a monotonic increase from $T^\ast$ down to $T_c$, showing no intervening decrease-in-$T$ behavior as depicted in Figs. \ref{fig-Deriv}(e) and \ref{fig-Deriv}(f). In this case, we note that, for $p\leq0.148$, a kink could be identified in a magnified view of the d$\rho$/d$T$ curve at a temperature above $T_c$, below which the increase of d$\rho$/d$T$ becomes much faster with decreasing temperature. This kink temperature is defined as the onset of superconducting fluctuations $T_f$ for these relatively high-doping samples, as shown in Fig. \ref{fig-Deriv}(h) for $p=0.145$. For the highest doping sample of $p=0.152$, however, we found that the increase of d$\rho$/d$T$ becomes rather smooth and shows no discernable kink features as temperature goes from $T^\ast$ down to $T_c$, making the above definition of $T_f$ not applied any more for this sample.

\begin{figure*}
  \includegraphics[scale=0.62]{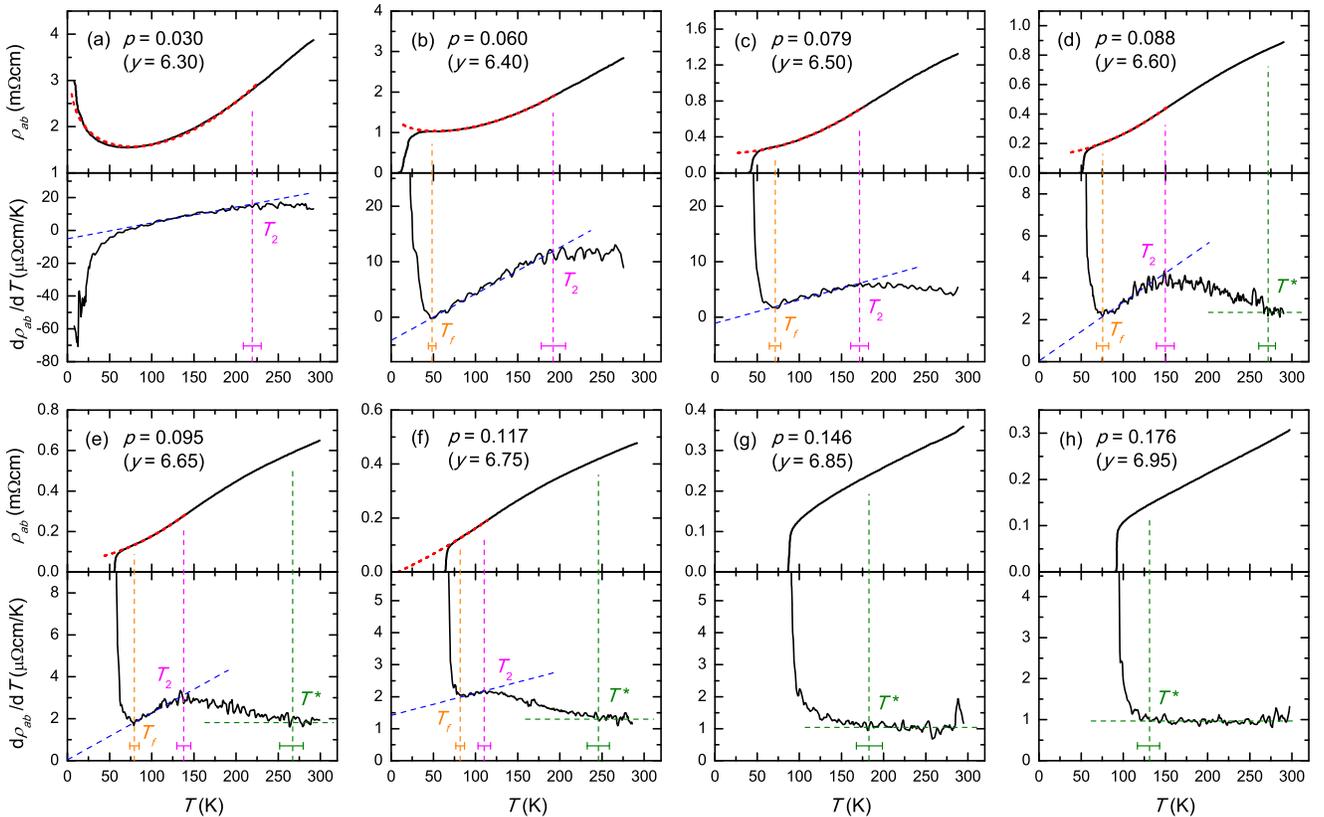}
  \caption{\label{fig-summary-thinfilm}(Color online) Doping evolution of the in-plane resistivity $\rho_{ab}(T)$ and its temperature derivative d$\rho_{ab}$/d$T$ in $c$-axis-oriented epitaxial thin films of YBa$_2$Cu$_3$O$_y$ (solid lines). The resistivity with oxygen content $y$ indicated is from Refs. \onlinecite{Wuyts1993,Wuyts1996}. The doping $p$ is determined from the $T_c$ of the films (following Ref. \onlinecite{Liang2006}), except for the non-superconducting film of $y=6.30$ shown in (a) for which the $p$ has been estimated based on the Hall coefficient measurement. \cite{Wuyts1996,Segawa2004} Three characteristic temperatures $T_f$, $T_2$, and $T^\ast$ are defined as explained in Fig. \ref{fig-Deriv}, with locations marked by orange, magenta, and olive vertical dashed lines, respectively. The horizontal error bars reflect the uncertainty in locating these temperatures. Blue diagonal dashed lines [(a)-(f), lower panels] and olive horizontal dashed lines [(d)-(h), lower panels] are guides to the eye, illustrating the $T$-linear behavior of d$\rho_{ab}$/d$T$ below $T_2$ and its $T$-constant behavior above $T^\ast$, respectively. The red dashed lines [(a)-(f), upper panels] are fits of $\rho_{ab}(T)$ to Eqs. (\ref{eq-doping-regime-1})-(\ref{eq-doping-regime-3}) in the range of temperature below $T_2$ (a) or between $T_2$ and $T_f$ [(b)-(f)]. See the text for details.}
\end{figure*}

\begin{figure*}
  \includegraphics[scale=0.62]{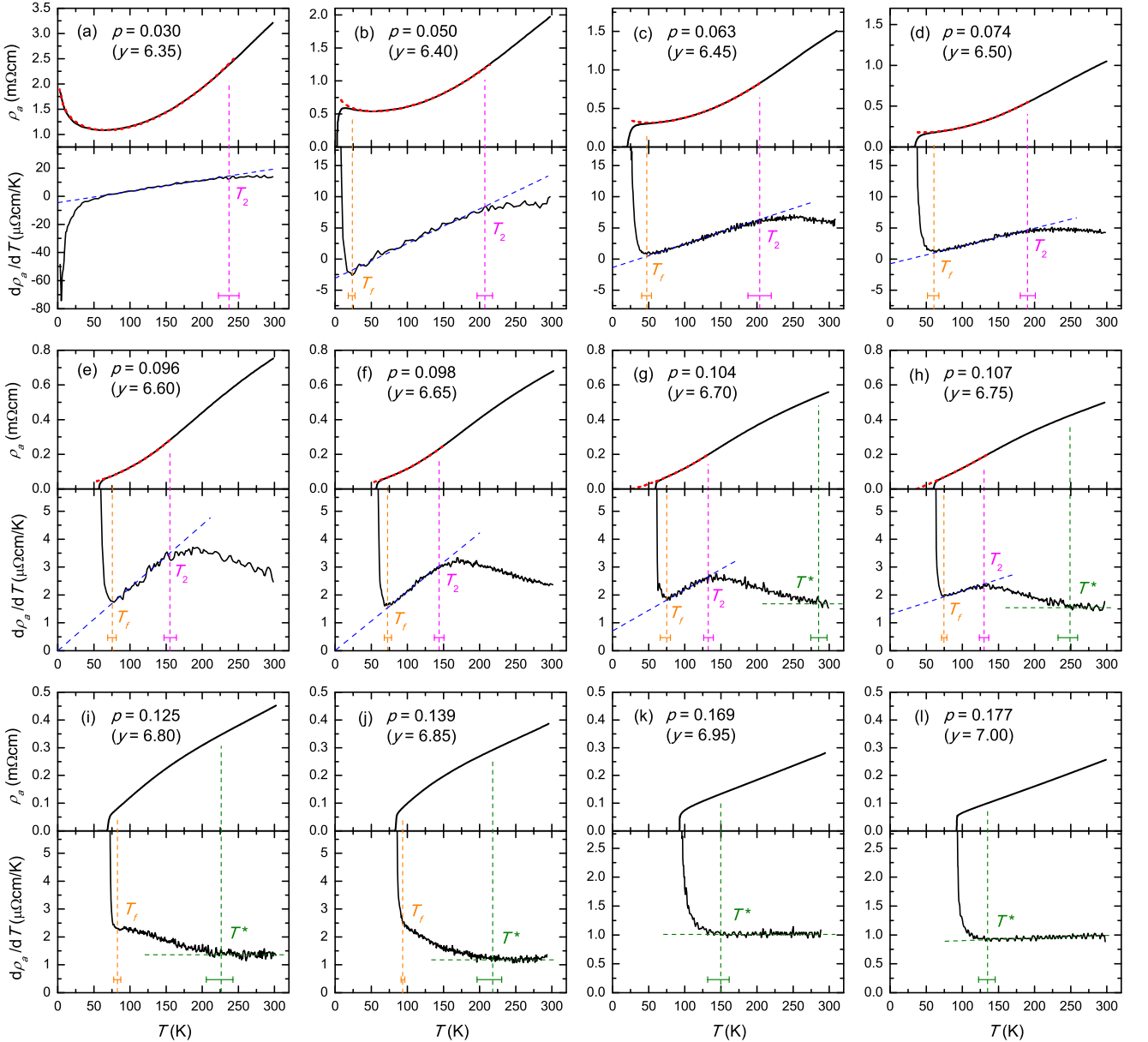}
  \caption{\label{fig-summary-single-crystal}(Color online) Doping evolution of the $a$-axis resistivity $\rho_{a}(T)$ and its temperature derivative d$\rho_{a}$/d$T$ in YBa$_2$Cu$_3$O$_y$ single crystals (solid lines). The resistivity with oxygen content $y$ indicated is from Refs. \onlinecite{Segawa2001,Ando2002a,Ando2002b}. The doping $p$ is determined from the $T_c$ of the crystals (following Ref. \onlinecite{Liang2006}), except for the non-superconducting crystal of $y=6.35$ shown in (a) for which the $p$ has been estimated based on the Hall coefficient measurement. \cite{Segawa2004} Three characteristic temperatures $T_f$, $T_2$, and $T^\ast$ are defined as explained in Fig. \ref{fig-Deriv}, with locations indicated by orange, magenta, and olive vertical dashed lines, respectively. The horizontal error bars reflect the uncertainty in locating these temperatures. Blue diagonal dashed lines [(a)-(h), lower panels] and olive horizontal dashed lines [(g)-(l), lower panels] are guides to the eye, highlighting the $T$-linear behavior of d$\rho_{a}$/d$T$ below $T_2$ and its $T$-constant behavior above $T^\ast$, respectively. The red dashed lines [(a)-(h), upper panels] are fits of $\rho_{a}(T)$ to Eqs. (\ref{eq-doping-regime-1})-(\ref{eq-doping-regime-3}) in temperature range below $T_2$ (a) or between $T_2$ and $T_f$ [(b)-(h)]. See the text for details.}
\end{figure*}

\begin{figure*}
  \includegraphics[scale=0.56]{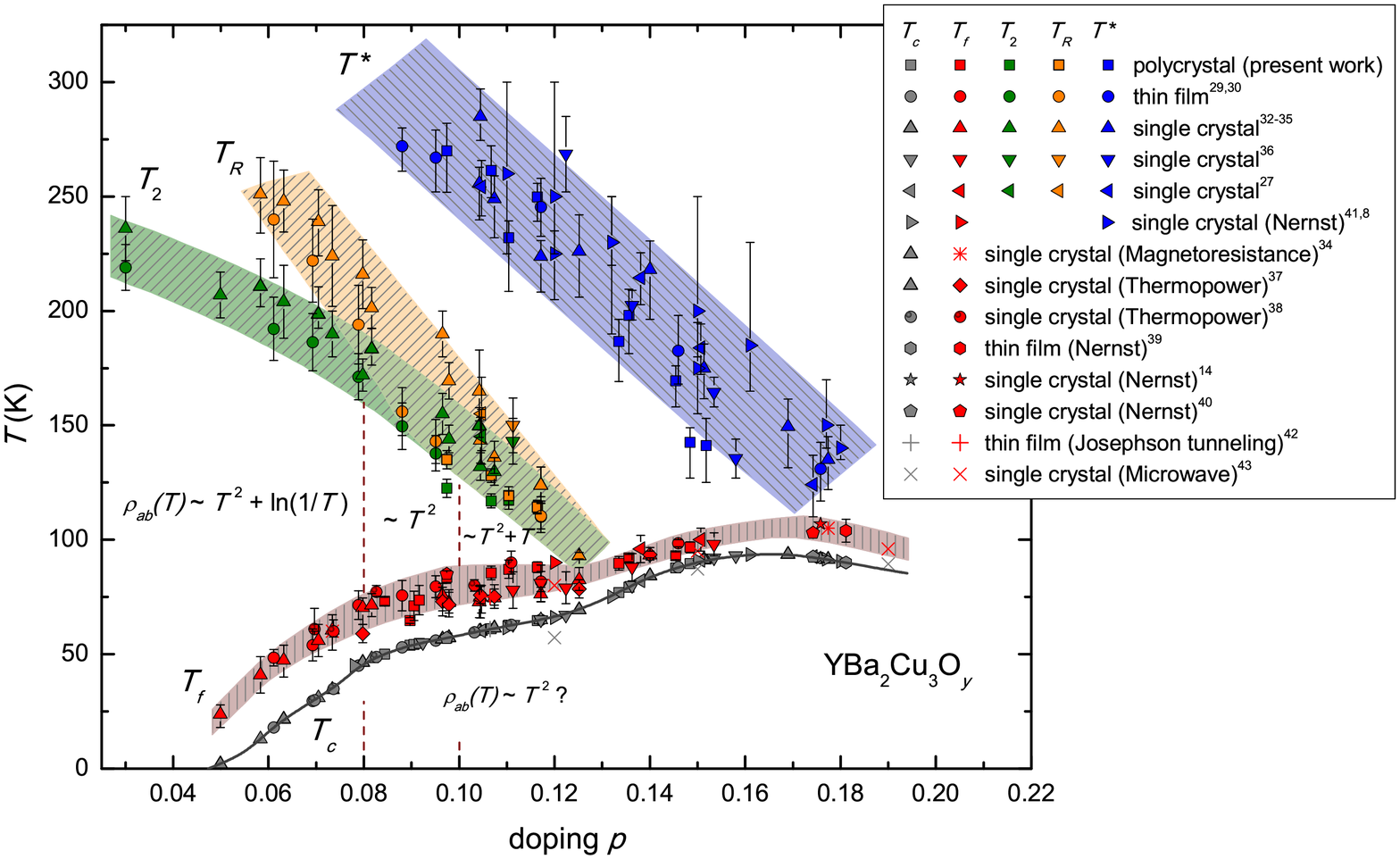}
  \caption{\label{fig-Phasediagram}(Color online) Temperature versus doping phase diagram of YBa$_2$Cu$_3$O$_y$ summarizing different temperature scales extracted from resistivity measurements. The labels $T_c$, $T_f$, $T_2$, $T_R$, and $T^\ast$, as explained in Figs. \ref{fig-Deriv} to \ref{fig-summary-single-crystal}, \ref{fig-summary-single-crystal-B}, and \ref{fig-summary-Res-and-Hall}, represent the superconducting transition temperature, the onset temperature for superconducting fluctuations, the temperature below which a predominant $T^2$ dependence emerges in $\rho_{ab}$, the temperature where $\rho_{ab}$ shows an inflection point (curvature change), and the pseudogap temperature, respectively. The results obtained from measurements on polycrystals, thin films, \cite{Wuyts1993,Wuyts1996} and single crystals \cite{Segawa2001,Ando2002a,Ando2002b,Lee2005,Ito1993,Takenaka1994} have been plotted together to demonstrate the consistency among different sample forms. Included are also the temperature $T_f$ determined from the magnetoresistance \cite{Ando2002b} and thermopower \cite{Segawa2003,Wang2001} measurements, as explained in the main text. For comparison, the $T_f$ identified in experiments of the Nernst effect, \cite{Xu2005,Wang2006,RullierAlbenque2006,Chang2010} Josephson effect, \cite{Bergeal2008} and microwave absorption, \cite{Grbic2011} as well as the pseudogap temperature $T^\ast$ defined from Nernst effect, \cite{Daou2010} are also shown. Error bars where present reflect the uncertainty in the determination of these temperature scales, as shown in Figs. \ref{fig-Deriv} to \ref{fig-summary-single-crystal}, \ref{fig-summary-single-crystal-B}, and \ref{fig-summary-Res-and-Hall}, or quoted from the references cited. \cite{Xu2005,Daou2010} Shaded areas are guides to the eye. The doping $p$ has been evaluated from the $T_c$ of the samples following Ref. \onlinecite{Liang2006} (gray solid line), or quoted from the references cited. \cite{Daou2010,Chang2010,Grbic2011} Two vertical dashed lines at $p=0.08$ and 0.10 divide the $T-p$ region below $T_2$ (and above $T_f$) into three parts to show that, besides a $T^2$-dependent component, the $\rho_{ab}(T)$ also involves a $\ln(1/T)$-dependent component at $p<0.08$ or, in the majority of samples, a $T$-linear component at $p>0.10$, as indicated in the figure. See the text for details.}
\end{figure*}

Here, we have described the doping evolution of $\rho(T)$ and its temperature derivative d$\rho$/d$T$ in YBCO polycrystals and defined three characteristic temperatures $T_f$, $T_2$, and $T^\ast$. It is necessary to check whether they really represent the in-plane transport properties of YBCO as assumed. For this purpose, we have re-examined typical results of the in-plane resistivity $\rho_{ab}(T)$ obtained on YBCO epitaxial thin films and single crystals, as demonstrated in Figs. \ref{fig-summary-thinfilm} and \ref{fig-summary-single-crystal} respectively. A set of $\rho_{ab}(T)$ data measured on $c$-axis-oriented epitaxial thin films in a wide range of doping (from Refs. \onlinecite{Wuyts1993,Wuyts1996}) and the corresponding temperature derivative d$\rho_{ab}$/d$T$ are shown in Fig. \ref{fig-summary-thinfilm} as solid lines. For the lightly-doped, non-superconducting film of $p=0.030$, Fig. \ref{fig-summary-thinfilm}(a) shows that the $\rho_{ab}(T)$ curve develops an upturn as temperature falls below about 75 K, making the temperature derivative d$\rho_{ab}$/d$T$ become negative and decrease rapidly. As temperature goes up from about 75 K, the d$\rho_{ab}$/d$T$ exhibits in approximate a $T$-linear increase until the temperature $T_2$ denoted is reached, suggesting a predominant $T^2$ dependence of $\rho_{ab}(T)$ in this temperature range. For superconducting films with doping $p$ from 0.060 to 0.117, Figs. \ref{fig-summary-thinfilm}(b) to \ref{fig-summary-thinfilm}(f) show that the d$\rho_{ab}$/d$T$ evolves with doping but at the same time remains similar in the overall, non-monotonic temperature dependence, as that shown in Figs. \ref{fig-Deriv}(e) and \ref{fig-Deriv}(f) for polycrystals at intermediate dopings. This has then enabled us to define temperature scales $T_f$, $T_2$, or $T^\ast$ for these films as described earlier for polycrystals. When doping is further increased, Figs. \ref{fig-summary-thinfilm}(g) and \ref{fig-summary-thinfilm}(h) show that the d$\rho_{ab}$/d$T$ changes to behave similar to that shown in Figs. \ref{fig-Deriv}(h) and \ref{fig-Deriv}(i) for polycrystals at relatively high dopings, i.e., increases monotonically as temperature decreases from $T^\ast$ down to $T_c$. Thus, through the above comparisons, we can see that, when doping grows from the point $p=0.097$, the $\rho(T)$ and d$\rho$/d$T$ for polycrystals shown in Figs. \ref{fig-RT} and \ref{fig-Deriv}(d) to \ref{fig-Deriv}(i) display qualitatively similar features to the $\rho_{ab}(T)$ and d$\rho_{ab}$/d$T$ for thin films shown in Figs. \ref{fig-summary-thinfilm}(e) to \ref{fig-summary-thinfilm}(h) at comparable dopings, evidencing that the $\rho(T)$ of polycrystalline samples primarily reflects in-plane resistive transport properties of YBCO in this doping regime. For dopings below 0.097, the $\rho(T)$ depicted in Fig. \ref{fig-RT} for polycrystals shows an upturn above $T_c$, whereas such an upturn is absent in $\rho_{ab}(T)$ of superconducting thin films as demonstrated in Figs. \ref{fig-summary-thinfilm}(b) to \ref{fig-summary-thinfilm}(d), thereby making the d$\rho$/d$T$ shown in Figs. \ref{fig-Deriv}(b) and \ref{fig-Deriv}(c) for polycrystals qualitatively different from the d$\rho_{ab}$/d$T$ shown in Figs. \ref{fig-summary-thinfilm}(b) to \ref{fig-summary-thinfilm}(d) for thin films apart from the presence of a minimum at $T_f$ in both cases. This difference could be related to the higher disorder level in polycrystals owing to the imperfect grain alignment than in thin films, which might then render the low-$T$ upturn phenomenon of $\rho_{ab}(T)$ observed in lightly-doped thin films [Fig. \ref{fig-summary-thinfilm}(a)] persisting to higher dopings and temperatures in polycrystals. In view of this situation, below we have excluded the $\rho(T)$ data of polycrystals for $p<0.097$ in discussing the $T$-dependence of in-plane resistivity in YBCO and only used them to extract the temperature $T_f$ for this doping region.

To further confirm the above findings, in Fig. \ref{fig-summary-single-crystal} we have reproduced the systematic $a$-axis resistivity $\rho_{a}(T)$ obtained by Ando and co-workers on YBCO single crystals across a wide doping regime (from Refs. \onlinecite{Segawa2001,Ando2002a,Ando2002b}), together with the corresponding temperature derivative d$\rho_{a}$/d$T$. In YBCO, the Cu-O chains lie along the $b$-axis, and hence the measurement of $\rho_{a}(T)$ avoids the contribution of Cu-O chains and probes solely the conduction of CuO$_2$ planes, making $\rho_{a}(T)$ the most accurate representation of the in-plane resistivity in YBCO. By comparing Fig. \ref{fig-summary-single-crystal} with Fig. \ref{fig-summary-thinfilm} and based on the above descriptions, one may readily perceive that, with doping varying from lightly-doped through underdoped to overdoped regime, the evolution of $\rho_{a}(T)$ and d$\rho_{a}$/d$T$ in single crystals looks quite similar to that of the $\rho_{ab}(T)$ and d$\rho_{ab}$/d$T$ in thin films, allowing one to define similarly the temperatures $T_f$, $T_2$, or $T^\ast$ for single crystals as illustrated in Fig. \ref{fig-summary-single-crystal}. This implies that the $\rho_{ab}(T)$ measurements in thin films have not been complicated by the small contribution of Cu-O chains, like that reported in previous study for single crystals, \cite{Ito1993,ARemark} and it further suggests that the essential features of the in-plane resistivity in YBCO, as revealed by the doping evolution of its temperature derivative, have remained the same among different sample forms.

Based on the above results, a phase diagram of YBCO has been constructed in Fig. \ref{fig-Phasediagram} depicting doping dependence of the characteristic temperatures determined from the in-plane resistivity shown in Figs. \ref{fig-RT} to \ref{fig-summary-single-crystal}. Besides aforementioned $T_f$, $T_2$, and $T^\ast$, another temperature scale labeled as $T_R$ is also plotted, which marks the inflection point in $\rho_{ab}(T)$ at some dopings and is extracted to compare with the temperatures $T_\mathrm{H}$ and $T_\mathrm{S}$ (plotted in Fig. \ref{fig-various-T-comparison}) derived respectively from the in-plane Hall coefficient and thermopower measurements of YBCO. Details on the definition of $T_R$, $T_\mathrm{H}$, and $T_\mathrm{S}$ as well as related discussions will be given in Figs. \ref{fig-summary-Res-and-Hall} to \ref{fig-various-T-comparison} and Sec. \ref{subsec:Res-Hall-Seebeck}. It is seen from Fig. \ref{fig-Phasediagram} that all the temperatures identified display good agreement among the values obtained from polycrystals, thin films, and single crystals, affirming their robustness against the variation of sample forms. This implies that they represent the generic property of the in-plane resistive transport in YBCO. In Fig. \ref{fig-Phasediagram}, it is noted that the temperature $T_f$ identified from a variety of other types of measurements such as the Nernst effect has also been included. This is to help make a better comparison of the result from resistivity to that from other experimental probes, as addressed in detail below.

\subsection{\label{subsec:SC-fluc}Onset temperature for superconducting fluctuations}

As pointed out earlier, in recent high-field magneto-transport study of overdoped LSCO, \cite{Rourke2011} the onset temperature for superconducting fluctuations above $T_c$ has been investigated by monitoring the temperature variation of the field dependence of the transverse magnetoresistance and identified as the temperature at which the magnetoresistance starts to show a deviation from its standard behavior (quadratic field dependence) in the normal state at high temperatures. \cite{Cooper2009} It has been further observed that this onset temperature identified coincides well with the temperature $T_f$ at which the d$\rho_{ab}$/d$T$ curve develops an upturn, suggesting the upturn in d$\rho_{ab}$/d$T$ caused by the fluctuating superconductivity and the $T_f$ just its onset temperature. \cite{Rourke2011} This proposes a convenient method to determine the temperature range of superconducting fluctuations in high-$T_c$ cuprates by locating the $T_f$ in d$\rho_{ab}$/d$T$, \cite{Rourke2011} as we have used in the present study. It is noticed that this method has also been adopted to investigate the superconducting fluctuations in other types of superconducting materials like the iron-based superconductor. \cite{Rey2013}

In YBCO, as shown above, the onset of the upturn in d$\rho_{ab}$/d$T$ at $T_f$ is easy to identify for low- and intermediate-doping samples as for these samples the d$\rho_{ab}$/d$T$ above $T_f$ decreases with $T$ decreasing and hence the $T_f$ marks a minimum in d$\rho_{ab}$/d$T$, which is similar to the case in overdoped LSCO. \cite{Rourke2011} For relatively high-doping samples of $p$ between about 0.13 and 0.15, although the d$\rho_{ab}$/d$T$ changes to rise monotonically from $T^\ast$ down to $T_c$ without an intervening minimum, the $T_f$ could still be identified as a kink temperature in d$\rho_{ab}$/d$T$ below which the rise of d$\rho_{ab}$/d$T$ becomes much steeper. For samples with doping higher than about 0.15, however, the d$\rho_{ab}$/d$T$ begins to show a rather smooth and continuous increase as temperature falls from $T^\ast$ down to $T_c$ [for instance, see Figs. \ref{fig-summary-single-crystal}(k) and \ref{fig-summary-single-crystal}(l)], which have made it difficult to define a $T_f$ like that in lower doping samples.

In Fig. \ref{fig-Phasediagram}, it is shown that the $T_f$ extracted from resistivity decreases with doping decreasing from nearly optimally doped to strongly underdoped regime, following a doping dependence quite similar to that of the $T_c$. The interval between $T_f$ and $T_c$, albeit getting a little bit wider as doping decreases, is confined in the range between about 5 K and 30 K. This suggests that the temperature range for superconducting fluctuations remains small in YBCO across the underdoped region of the phase diagram. By contrast, Fig. \ref{fig-Phasediagram} shows that the identified pseudogap temperature $T^\ast$, which is in broad agreement with that determined from other measurements such as neutron scattering \cite{Fauque2006,Mook2008} and Nernst effect, \cite{Daou2010} increases roughly linearly to quite high values as doping decreases from the slightly overdoped to deeply underdoped regime, as typically reported in previous studies. \cite{Timusk1999,Norman2005} This therefore shows that the $T_f$ exhibits an opposite doping dependence to that of the $T^\ast$ for $p\lesssim0.15$. Combined with the much lower value of $T_f$ than the $T^\ast$, it suggests that in YBCO the region of superconducting fluctuations above $T_c$ occupies only a small portion of the pseudogap phase in the $T-p$ phase diagram, and that there seems no obvious correlation between the pseudogap opening and the presence of superconducting fluctuations.

In YBCO, we note that the transverse magnetoresistance along the $a$-axis $\Delta\rho_a(H)$ has been measured \cite{Ando2002b} on the same crystals as that used to measure the $a$-axis resistivity $\rho_a(T)$ as reproduced in Fig. \ref{fig-summary-single-crystal}, which provides a good opportunity to check the correspondence between the $T_f$ from resistivity and that from magnetoresistance as observed in overdoped LSCO. \cite{Rourke2011,BRemark} For the crystal of $y=6.50$ ($p=0.074$), it has been shown that the $\Delta\rho_a(H)$ recovers its typical $H^2$ dependence in the whole range of sweeping magnetic fields as temperature increases to 60 K [Fig. 1(d) of Ref. \onlinecite{Ando2002b}], suggesting the vanishing of superconducting fluctuations at this temperature according to the criterion used in the study of overdoped LSCO. \cite{Rourke2011} Note that this finding corresponds just with the $T_f$ of 60 K defined in resistivity for the same crystal, as indicated in Fig. \ref{fig-summary-single-crystal}(d), confirming the validity of extending the observation in LSCO to YBCO to identify the upturn in d$\rho_{ab}$/d$T$ with the onset of fluctuating superconductivity. \cite{CRemark} Furthermore, in the slightly overdoped crystal of $y=7.00$ ($p=0.177$), for which the $T_f$ is difficult to define in d$\rho_{ab}$/d$T$ as explained above, the magnetoresistance measurement indicates a $T_f$ of about 105 K, i.e., 13 K above $T_c$ [Fig. 1(c) of Ref. \onlinecite{Ando2002b}], showing that the superconducting fluctuations in YBCO survive to a restricted temperature range above $T_c$ in the overdoped regime as well.

It is noted that a small temperature range for superconducting fluctuations in YBCO has also been reported by using other experimental techniques. In Fig. \ref{fig-Phasediagram}, a collection of the $T_f$ derived from several Nernst effect, \cite{Xu2005,Wang2006,RullierAlbenque2006,Chang2010} Josephson junction, \cite{Bergeal2008} and microwave absorption \cite{Grbic2011} experiments has been shown. They are defined as the onset of a large Nernst signal, \cite{Xu2005,Wang2006,RullierAlbenque2006,Chang2010} the onset of an excess conductance peak due to the Josephson effect, \cite{Bergeal2008} or the onset of an excess microwave conductivity \cite{Grbic2011} as temperature decreases. In slightly overdoped regime, Fig. \ref{fig-Phasediagram} reveals that the $T_f$ yielded from these studies is about 7-15 K higher than $T_c$ and exhibits a doping dependence roughly parallel to that of the $T_c$, similar to the behavior of $T_f$ from resistivity in the underdoped regime. As doping decreases, Fig. \ref{fig-Phasediagram} illustrates that the $T_f$ from these studies agrees quite well with that from resistivity, showing values of about 8 K higher than $T_c$ at $p\simeq0.15$ and of about 15-28 K higher than $T_c$ for $p$ roughly between 0.10 and 0.12. This good agreement among various measurements, on the one hand, lends additional strong support to the proposal that the upturn in d$\rho_{ab}$/d$T$ signifies the appearance of superconducting fluctuations, and on the other hand, suggests further that the superconducting fluctuations in YBCO persist only to a narrow range of temperature above $T_c$, well below $T^\ast$, throughout nearly the whole doping phase diagram.

We should point out, however, that, contrary to the above results, a broad temperature range of superconducting fluctuations in YBCO has also been inferred from some other kinds of experiments, notably from experiments of torque magnetization \cite{Li2010} and infrared conductivity. \cite{Dubroka2011} In a slightly overdoped, twinned single crystal of $p\simeq0.17$ ($T_c=92$ K), an onset of diamagnetism at $T\sim130$ K, i.e., about 40 K above $T_c$ has been reported in torque magnetometry measurement. \cite{Li2010} That the observed diamagnetic signal is both strongly temperature-dependent and nonlinear in magnetic field has been taken as strong evidence for superconducting fluctuations as its origin. \cite{Li2010} As a consequence it is suggested that the fluctuating superconductivity in this sample extend to relatively high temperatures above $T_c$. \cite{Li2010} It is worth noting that the onset temperature $T_f$ of about 130 K is $\sim23$ K higher than the value of 107 K (15 K above $T_c$) from Nernst effect \cite{Wang2006} as mentioned earlier and shown in Fig. \ref{fig-Phasediagram}, which was obtained on the same doping sample. \cite{Li2010,Wang2006} This discrepancy in the results from torque magnetization and Nernst effect calls for further understanding. In a recent study of the infrared $c$-axis conductivity in YBCO, \cite{Dubroka2011} on the basis of a multilayer model analysis of the spectra to differentiate intra-bilayer and inter-bilayer conductivities, precursor superconductivity above $T_c$ has been suggested to underlie the observation of the development of a transverse plasma mode and related anomalies in certain infrared-active phonons. The onset temperature of the anomalous phonon softening or broadening, identified accordingly as the onset of precursor superconductivity, was shown \cite{Dubroka2011} to first increase roughly linearly from a value close to $T_c$ at slightly overdoped side to a high value of about 180 K with doping falling to $p\sim0.08-0.06$ and then decrease roughly linearly to a value of about 50 K with doping falling further to $p\simeq0.03$, as replotted in Fig. \ref{fig-various-T-comparison}(b) as $T_{\mathrm{infrared}}$. It is hence suggested that the superconducting fluctuations in YBCO survive to considerably high temperatures in most part of the underdoped regime, including the part of $0.03\lesssim p\lesssim0.05$ where bulk superconductivity is absent. \cite{Dubroka2011} Clearly, the onset of superconducting fluctuations deduced from this study is much higher and shows a distinct doping dependence compared to the $T_f$ from other studies as summarized in Fig. \ref{fig-Phasediagram}. The reason behind this contrast needs to be elucidated. In this regard, we note, interestingly, that the $T_{\mathrm{infrared}}$ line identified in the above infrared-conductivity study, in fact, coincides well with the $T_R$, $T_\mathrm{H}$, and $T_\mathrm{S}$ defined respectively in resistivity, Hall effect, and thermopower measurements for $p\gtrsim0.08$, as revealed in Fig. \ref{fig-various-T-comparison}(b), hinting at a possible connection among these characteristic temperatures. We shall come back to this point in Sec. \ref{subsec:Res-Hall-Seebeck}.

It is worth emphasizing that controversy on the temperature range of superconducting fluctuations, whether it is small or broad, has also emerged for other families of high-$T_c$ cuprates such as LSCO and Bi$_2$Sr$_2$CaCu$_2$O$_{8+\delta}$ (Bi2212). Unlike in YBCO, Nernst effect measurements have detected the onset of a large Nernst signal well above $T_c$ in LSCO and Bi2212 across the doping phase diagram. \cite{Wang2006} The onset temperature, almost 100 K or 80 K above $T_c$ at some underdoping point in LSCO or Bi2212 respectively, \cite{Wang2006} has been shown further to be in accordance with the onset of diamagnetism observed in torque magnetometry in these two cuprates, \cite{Wang2005,Li2010} indicating the presence of superconducting fluctuations in a rather broad temperature range of the pseudogap phase. On the other side, however, the a.c. conductivity at microwave \cite{Kitano2006} or terahertz \cite{Bilbro2011} frequencies in LSCO and that at terahertz frequencies \cite{Corson1999} in Bi2212 have all probed a much smaller temperature range of superconducting fluctuations, at most 16 K above $T_c$ in LSCO over the whole superconducting dome \cite{Kitano2006,Bilbro2011} and about 20 K above $T_c$ in an underdoped Bi2212 thin film. \cite{Corson1999} To reconcile these apparently conflicting results, different scenarios have been put forward, including, for instance, the attribution of the appearance of Nernst or diamagnetic signal well above $T_c$ to something other than superconducting fluctuations, \cite{Cyr-Choiniere2009} or the possibility of fluctuating superconductivity in high-$T_c$ cuprates having unusual properties to make it manifested in distinct manners in different kinds of experiments. \cite{Bilbro2011PRB} In comparing the $T_f$ determined for different cuprate families but from the same experimental method such as the Nernst effect measurement, it was ever suggested that intrinsic disorder of the sample may play a role in governing the temperature range of superconducting fluctuations, \cite{RullierAlbenque2006} which might explain the rather narrower fluctuation regime in YBCO than in LSCO as YBCO is widely recognized to be a clean material while in LSCO the intrinsic disorder should be considerably higher. To clarify the above issues and to eventually resolve the debate on the temperature extension of superconducting fluctuations in high-$T_c$ cuprates, more experiment and analysis need to be carried out.

\subsection{\label{subsec:T-square-Res}Doping variation of $T^2$-dependent resistivity in the pseudogap phase}

Now let us turn to the temperature regime above $T_f$. In the $T-p$ phase diagram of Fig. \ref{fig-Phasediagram}, it is shown that, at temperatures about half the $T^\ast$, there lies the characteristic temperature $T_2$, which decreases roughly in parallel with $T^\ast$ as doping increases and finally meets the temperature $T_f$ at $p\sim0.13$, thereby dividing the regime between $T^\ast$ and $T_f$ into two regions. Because the $T_2$ marks the temperature below which the d$\rho_{ab}$/d$T$ exhibits a linear-$T$ dependence, it shows that in the lower temperature region between $T_2$ and $T_f$, the $\rho_{ab}(T)$ develops a $T^2$ dependence, while in the higher temperature region between $T^\ast$ and $T_2$, a gradual transition of $\rho_{ab}(T)$ from the $T$-linear behavior above $T^\ast$ to the $T^2$ variation below $T_2$ takes place. It is noted that the overall temperature evolution of $\rho_{ab}(T)$ within the pseudogap phase has been qualitatively referred to as an ``$S$-shape'' character \cite{Hussey2008} of the $\rho_{ab}(T)$, for instance, by Ando \textit{et al.} in their study of the second derivative of normalized in-plane resistivity d$^2$$\rho_{ab}^n$/d$T^2$ [$\rho_{ab}^{n}=\rho_{ab}/\rho_{ab}(300~\mathrm{K})$]. \cite{Ando2004} Subsequently, the emergence of a $T^2$-dependent $\rho_{ab}(T)$ in underdoped YBCO has been pointed out. \cite{Lee2005,Rullier-Albenque2007,LeBoeuf2011} More recently, the presence of $T^2$-dependent resistivity in the pseudogap phase has also been noted by Bari{\v s}i{\'c} \textit{et al.} in investigating the $\rho_{ab}(T)$ in HgBa$_2$CuO$_{4+\delta}$ (Hg1201) and suggested as a common feature across different families of high-$T_c$ cuprates. \cite{Barisic2013} By examining the first derivative of in-plane resistivity d$\rho_{ab}$/d$T$, we have been able to extract systematically the doping variation of the temperature boundary $T_2$ in YBCO as depicted in Fig. \ref{fig-Phasediagram} to map out the $T-p$ region over which the $\rho_{ab}(T)$ contains a $T^2$-dependent component. Below it is shown that more detailed information concerning this particular $T-p$ region in YBCO could be obtained by taking a closer look at the doping evolution of d$\rho_{ab}$/d$T$.

\begin{figure}
  \includegraphics[scale=0.3]{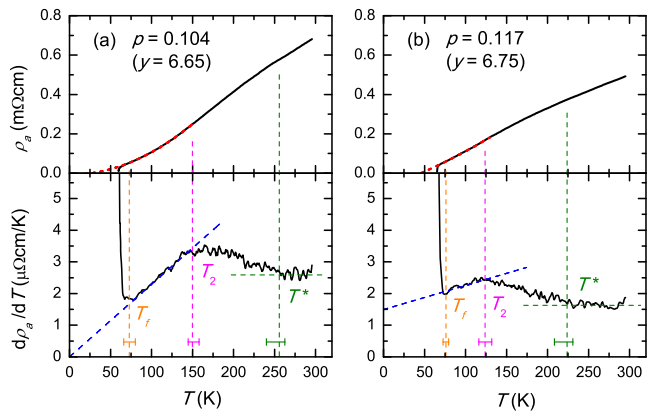}
  \caption{\label{fig-summary-single-crystal-B}(Color online) The $a$-axis resistivity $\rho_{a}(T)$ and its temperature derivative d$\rho_{a}$/d$T$ for YBa$_2$Cu$_3$O$_y$ single crystals (solid lines) at two dopings $p=0.104$ (a) and 0.117 (b). The $\rho_{a}(T)$ with oxygen content $y$ indicated is from Ref. \onlinecite{Lee2005} and the doping $p$ is determined from the $T_c$ of the crystals following Ref. \onlinecite{Liang2006}. The three temperatures $T_f$, $T_2$, and $T^\ast$, and the fits to $\rho_{a}(T)$ between $T_f$ and $T_2$ are illustrated as in Fig. \ref{fig-summary-single-crystal}. In (a), the $\rho_{a}(T)$ is fitted to Eq. (\ref{eq-doping-regime-2}), as the linear d$\rho_{a}$/d$T$ shows zero intercept at $T=0$ K, while in (b), the $\rho_{a}(T)$ is fitted to Eq. (\ref{eq-doping-regime-3}) in accordance with a positive intercept of the d$\rho_{a}$/d$T$. See the text for details.}
\end{figure}

For both thin films and single crystals, Figs. \ref{fig-summary-thinfilm} and \ref{fig-summary-single-crystal} show that, as doping varies, the $T$-linear dependence of d$\rho_{ab}$/d$T$ below $T_2$ exhibits varying intercept on the vertical axis at zero temperature, as indicated by blue diagonal dashed lines in both figures. In Fig. \ref{fig-summary-thinfilm}, we can see, for dopings less than 0.08 [Figs. \ref{fig-summary-thinfilm}(a) to \ref{fig-summary-thinfilm}(c)], the intercept shows negative values; as doping increases to be between 0.08 and 0.10 [Figs. \ref{fig-summary-thinfilm}(d) and \ref{fig-summary-thinfilm}(e)], the intercept becomes zero; with doping further increasing [Fig. \ref{fig-summary-thinfilm}(f)], the intercept changes to be positive. In Fig. \ref{fig-summary-single-crystal} for single crystals, this doping evolution of the intercept has been well repeated: for $p<0.08$ [Figs. \ref{fig-summary-single-crystal}(a) to \ref{fig-summary-single-crystal}(d)], the intercept is negative; when $p$ is about 0.10 [Figs. \ref{fig-summary-single-crystal}(e) and \ref{fig-summary-single-crystal}(f)], the intercept is zero; while for $p>0.10$ [Figs. \ref{fig-summary-single-crystal}(g) and \ref{fig-summary-single-crystal}(h)], the intercept has a positive value. We note that, in earlier reports of the $\rho_{a}(T)$ in single crystals,\cite{Ito1993,Takenaka1994} the derivative d$\rho_{a}$/d$T$ shows agreement with this trend. In the $\rho_{a}(T)$ report of Ref. \onlinecite{Lee2005}, the YBCO single crystals were also grown by Ando \textit{et al.} and the d$\rho_{a}$/d$T$ exhibits essentially the same doping evolution as that shown in Fig. \ref{fig-summary-single-crystal}, except for $p=0.104$ ($y=6.65$) at which, as shown in Fig. \ref{fig-summary-single-crystal-B}(a), the d$\rho_{a}$/d$T$ between $T_2$ and $T_f$ has zero intercept at $T=0$ K, different from the positive intercept of d$\rho_{a}$/d$T$ shown in Fig. \ref{fig-summary-single-crystal}(g) at the same doping level ($T_c=60$ K for both crystals). It is also noted that, the $\rho_{a}(T)$ below 120 K is shown for a YBCO single crystal of $p=0.11$ in Ref. \onlinecite{LeBoeuf2011} and the approximate $T$-linear d$\rho_{a}$/d$T$ between 120 K and $T_f\simeq72$ K shows zero intercept at $T=0$ K [see Fig. \ref{fig-Rho-in-field-comparison}(b)]. By summarizing the results as outlined above, we may see that, for dopings in the window of $0.08\lesssim p\lesssim0.10$, the linear d$\rho_{ab}$/d$T$ in thin films and single crystals of YBCO always shows zero intercept at zero temperature, implying a pure $T^2$-dependent $\rho_{ab}(T)$ in the temperature range between $T_2$ and $T_f$. For $0.10<p\lesssim0.11$, the $T=0$ intercept of d$\rho_{ab}$/d$T$ may be zero in some crystals, implying similarly a pure $T^2$-dependent $\rho_{ab}(T)$ in these samples, while it may also be positive in other crystals, implying the presence of an additional contribution to $\rho_{ab}(T)$ aside from the $T^2$-dependent component in such samples. The indication of additional contribution to the $\rho_{ab}(T)$ at $T_f<T<T_2$ also shows for $p<0.08$ where the intercept of d$\rho_{ab}$/d$T$ is negative and for $0.11<p<0.13$ where the intercept is positive.

It is natural to consider the presence of an additional $T$-linear term in $\rho_{ab}(T)$ to account for the positive intercept of d$\rho_{ab}$/d$T$ at zero temperature. That is, in samples of $0.10<p<0.13$ which show positive intercept of the d$\rho_{ab}$/d$T$, the $\rho_{ab}(T)$ involves a $T$-linear component besides the $T^2$-dependent one at temperatures between $T_2$ and $T_f$. For $p<0.08$, $\rho_{ab}(T)$ and d$\rho_{ab}$/d$T$ of the lightly-doped, non-superconducting sample of $p=0.03$ [Figs. \ref{fig-summary-thinfilm}(a) and \ref{fig-summary-single-crystal}(a)] suggest that, in superconducting samples, the additional contribution to the $\rho_{ab}(T)$ between $T_2$ and $T_f$ may come from the small high-temperature tail of a low-$T$ upturn in resistivity, which leads to the intercept of d$\rho_{ab}$/d$T$ to be negative. This could be perceived in particular in Fig. \ref{fig-summary-single-crystal}(b) for $p=0.05$ at which the low-$T$ upturn in resistivity has already shown up at $T>T_f$. One may note that the above detailed doping evolution of $\rho_{ab}(T)$, as indicated from the d$\rho_{ab}$/d$T$, has been concealed in the plot of second derivative d$^2$$\rho_{ab}^n$/d$T^2$ in previous study. \cite{Ando2004}

\begin{figure}
  \includegraphics[scale=0.3]{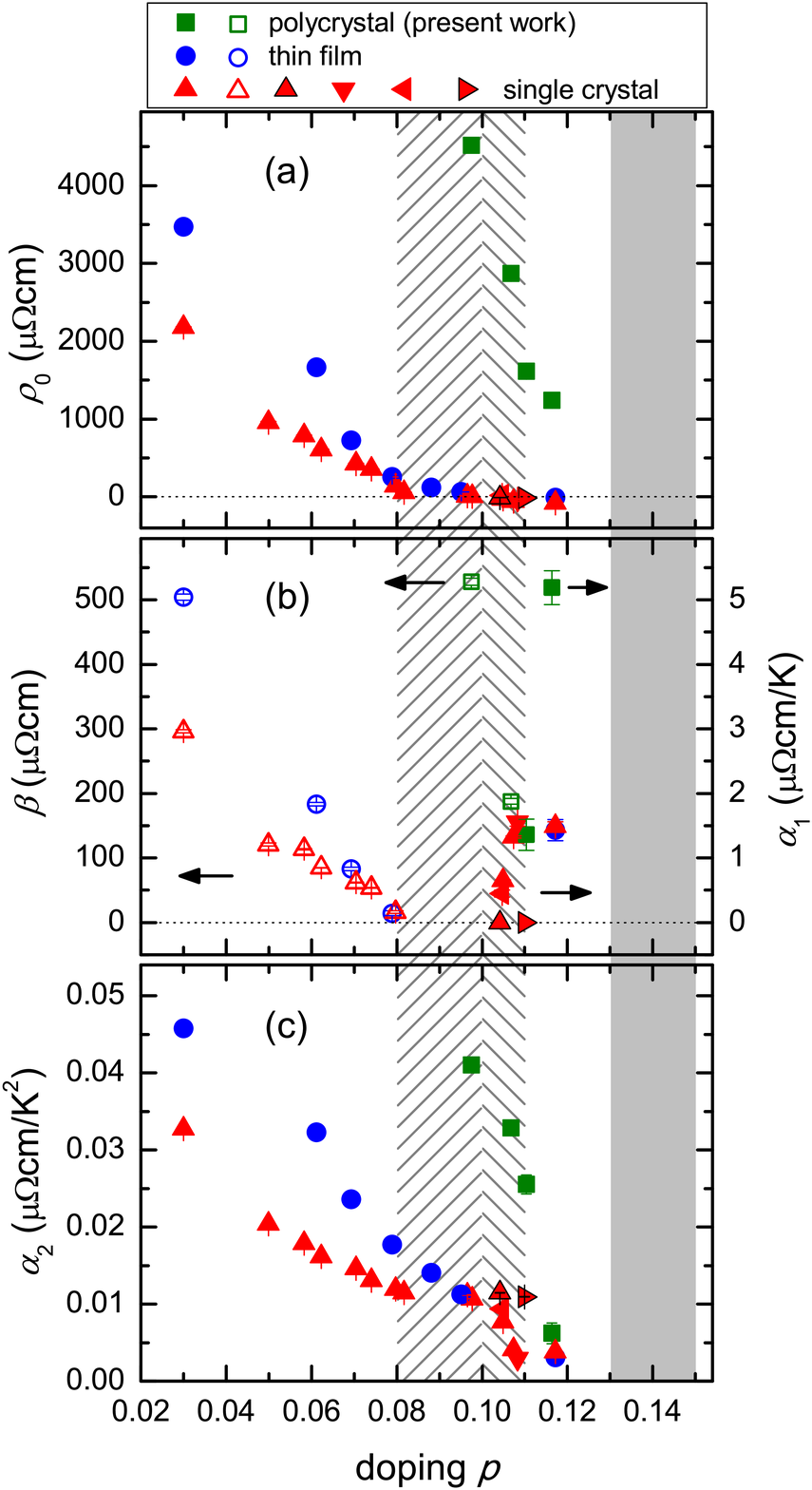}
  \caption{\label{fig-summary-fit-coefficients}(Color online) Doping evolution of the coefficients obtained in fitting the resistivity of YBa$_2$Cu$_3$O$_y$ shown in Figs. \ref{fig-RT}, \ref{fig-summary-thinfilm}, \ref{fig-summary-single-crystal}, and \ref{fig-summary-single-crystal-B} to Eqs. (\ref{eq-doping-regime-1})-(\ref{eq-doping-regime-3}). The fits, shown in Figs. \ref{fig-RT}, \ref{fig-summary-thinfilm}, \ref{fig-summary-single-crystal}, and \ref{fig-summary-single-crystal-B} as red dashed lines, are conducted for $T<T_2$ in non-superconducting samples or $T_f<T<T_2$ in superconducting samples. The results in fitting to the $\rho_{a}(T)$ of single crystals from Refs. \onlinecite{Ito1993,Takenaka1994} (not shown) and Ref. \onlinecite{LeBoeuf2011} [shown in Fig. \ref{fig-Rho-in-field-comparison}(b)] are also included (down, left, and right triangles, respectively). The doping regime above $p=0.13$ is painted gray to indicate that the fits could be performed only for samples of $p<0.13$ because the temperature $T_2$ declines to meet the $T_f$ at around this doping level (see Fig. \ref{fig-Phasediagram}). The hatching at $0.08\leq p\leq0.10$ is to indicate the doping region at which the $\rho_{ab}(T)$ in thin films and single crystals shows a pure $T^2$ dependence at $T_f<T<T_2$. The hatching at $p$ between 0.10 and 0.11 is to indicate the doping interval over which the $\rho_{a}(T)$ may still follow a pure $T^2$ dependence in some single crystals (red triangles with black edges), or show a coexistence of the $T$-linear and $T^2$-dependent terms (nonzero $\alpha_1$) in many other crystals. See the text for details.}
\end{figure}

To crosscheck the above findings, we have made a quantitative fit to $\rho_{ab}(T)$ in the temperature range between $T_2$ and $T_f$. Previously, it has been observed that, by using high magnetic fields to strip away superconductivity, the exposed normal-state resistivity shows a low-$T$ upturn and follows a $\ln(1/T)$ dependence in several underdoped high-$T_c$ compounds such as LSCO and Bi$_2$Sr$_{2-x}$La$_x$CuO$_{6+\delta}$. \cite{Ando1995,Boebinger1996,Ono2000} In YBCO, it has also been pointed out that, for lightly-doped, non-superconducting single crystals, the low-$T$ upturn of the $\rho_{ab}(T)$ could be well described by the $\ln(1/T)$ dependence. \cite{Sun2005,Doiron-Leyraud2006} In view of these, we have attempted to fit the $\rho_{ab}(T)$ in YBCO thin films or single crystals to
\begin{equation}
\rho_{ab}(T)=\rho_{0}+\alpha_{2}T^2+\beta\ln(1/T)\label{eq-doping-regime-1}
\end{equation}
for $p<0.08$ at temperatures below $T_2$ (at $p=0.03$) or between $T_2$ and $T_f$. Here, $\rho_{0}$ represents the residual resistivity at $T=0$~K, and $\alpha_{2}$ and $\beta$ are the coefficients of the $T^2$- and $\ln(1/T)$-dependent terms, respectively. For samples of $0.08\lesssim p\lesssim0.10$ and the samples showing zero intercept of d$\rho_{ab}$/d$T$ with $0.10<p\lesssim0.11$, the $\rho_{ab}(T)$ between $T_2$ and $T_f$ is fitted to
\begin{equation}
\rho_{ab}(T)=\rho_{0}+\alpha_{2}T^2,\label{eq-doping-regime-2}
\end{equation}
while for samples of $0.10<p<0.13$ showing the positive intercept of d$\rho_{ab}$/d$T$, the $\rho_{ab}(T)$ between $T_2$ and $T_f$ is fitted to
\begin{equation}
\rho_{ab}(T)=\rho_{0}+\alpha_{2}T^2+\alpha_{1}T,\label{eq-doping-regime-3}
\end{equation}
where $\alpha_{1}$ is the coefficient of the $T$-linear term.

The above fits have been displayed as red dashed lines in Figs. \ref{fig-summary-thinfilm}, \ref{fig-summary-single-crystal}, \ref{fig-summary-single-crystal-B}, and \ref{fig-Rho-in-field-comparison}, where we could see they reproduce quite well the $\rho_{ab}(T)$ data between $T_2$ and $T_f$, corroborating the indications from d$\rho_{ab}$/d$T$. In Fig. \ref{fig-summary-fit-coefficients}, the obtained fitting coefficients have been summarized. It is shown that, for both thin films and single crystals, the residual resistivity $\rho_{0}$ [Fig. \ref{fig-summary-fit-coefficients}(a)] decreases rapidly with increasing doping and becomes negligible when $p$ is above 0.08. The coefficient $\beta$ [Fig. \ref{fig-summary-fit-coefficients}(b), left-hand axis] shows similarly a rapid decline with doping, suggesting a fadeaway of the low-$T$ upturn in resistivity as doping rises and accordingly a metal-insulator crossover at $p\simeq0.08$. It is helpful to note that this is in line with the observation in heat transport that the thermal conductivity of YBCO at very low temperature changes from a decrease-in-magnetic-field to increase-in-magnetic-field behavior as doping rises across $p\simeq0.08$, indicating as well a metal-insulator crossover at about the same doping. \cite{Sun2004} Recently, this phenomenon of metal-insulator crossover has been suggested to arise from a change in the Fermi-surface topology of YBCO at $p\simeq0.08$, which is in turn indicated from the observation that the Hall coefficient $R_\mathrm{H}$ of YBCO in high magnetic fields exhibits a qualitative change in its temperature dependence as doping passes through the point of 0.08. \cite{LeBoeuf2011} Note that, in the vicinity of $p\simeq0.08$, critical changes or anomalies in other physical properties of YBCO have also been reported; for instance, a collapse of the spin density wave phase \cite{Coneri2010,Haug2010} or a divergence in the quasiparticle effective mass \cite{Sebastian2010} have been detected as this doping level is approached from below or above respectively.

When doping rises to be between about 0.08 and 0.10, the well description of $\rho_{ab}(T)$ by Eq. (\ref{eq-doping-regime-2}) confirms its pure $T^2$ dependence for temperatures between $T_2$ and $T_f$ in both thin films and single crystals. As doping further proceeds, to the $\rho_{ab}(T)$ data showing positive intercept in d$\rho_{ab}$/d$T$, it becomes necessary to include the $T$-linear term, namely, to use Eq. (\ref{eq-doping-regime-3}) to obtain a satisfactory fit. In this case, Fig. \ref{fig-summary-fit-coefficients}(b) (right-hand axis) shows that the coefficient $\alpha_1$ increases with increasing doping, suggesting a growing strength of the $T$-linear term. In Fig. \ref{fig-summary-fit-coefficients}(c), it is shown that, similar to the report in previous study, \cite{Barisic2013} the coefficient $\alpha_2$ of the $T^2$ term decreases roughly continuously as doping increases. At $0.10<p<0.13$, the $\alpha_1$ and $\alpha_2$ exhibit opposite doping dependences when they both appear. In crystals of $0.10<p\lesssim0.11$ showing no $T$-linear term in $\rho_{ab}(T)$, the $\alpha_2$ [red triangles with black edges in Fig. \ref{fig-summary-fit-coefficients}(c)] shows values close to that at $0.08\lesssim p\lesssim0.10$. In Fig. \ref{fig-summary-fit-coefficients}, the coefficients yielded in fitting the resistivity of polycrystals have also been plotted. The fits, shown in Fig. \ref{fig-RT} as red dashed lines, are conducted for $p\geq0.097$ samples at temperatures between $T_2$ and $T_f$. As indicated in Fig. \ref{fig-summary-fit-coefficients}(b), compared to thin films and single crystals, in polycrystals the metal-insulator crossover in $\rho(T)$ seems to locate at higher doping point $p\simeq0.11$ and there seems to be no doping interval for $\rho(T)$ to exhibit a pure $T^2$ dependence. In addition, it is shown in Fig. \ref{fig-summary-fit-coefficients} that the fitting coefficients such as the $\alpha_1$ and $\alpha_2$ have considerably higher values in polycrystals. These features should be related to, as aforementioned, the involvement of $c$-axis contribution in $\rho(T)$ and the substantially higher level of disorder in polycrystals, as evidenced by the quite large $\rho_0$ shown in Fig. \ref{fig-summary-fit-coefficients}(a). Despite this, it is worth emphasizing that the temperature $T_2$ in polycrystals shows essentially the same values as that in thin films and single crystals (Fig. \ref{fig-Phasediagram}), illustrating that the onset of the $T^2$-dependent resistivity in the pseudogap phase is an intrinsic property of YBCO determined predominantly by the hole doping concentration in the CuO$_2$ planes.

\begin{figure}
  \includegraphics[scale=0.3]{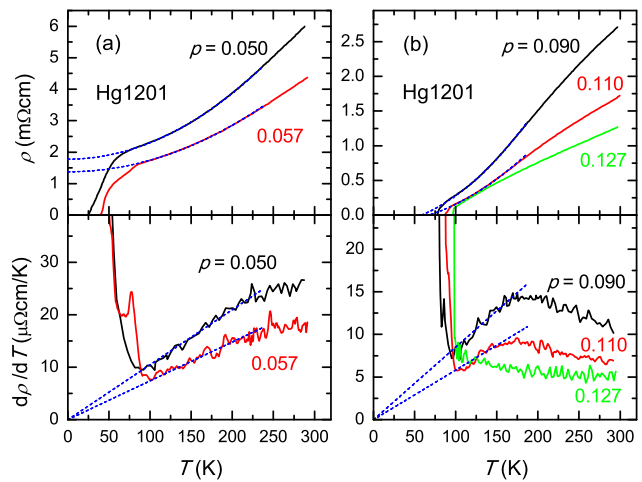}
  \caption{\label{fig-Hg120-RT-underdoped}(Color online) $\rho(T)$ and the corresponding temperature derivative d$\rho$/d$T$ of underdoped HgBa$_2$CuO$_{4+\delta}$ (Hg1201) polycrystalline samples. The $\rho(T)$ data with $p$ ranging from 0.050 to 0.127 are taken from Ref. \onlinecite{Yamamoto2000}. Blue dashed lines are fits of the resistivity ($p\leq0.110$) to $\rho(T)=\rho_{0}+\alpha_{2}T^2$, that is, Eq. \ref{eq-doping-regime-2}, which is shown to describe well the data at intermediate temperatures. One difference between (a) and (b) is that in (a) the obtained $\rho_{0}>0$ while in (b) $\rho_{0}<0$, as can be perceived by the intercepts of the blue dashed lines on the vertical or horizontal axis in the upper panels.}
\end{figure}

It is instructive to compare the doping evolution of $\rho_{ab}(T)$ below $T_2$ as uncovered above in YBCO with that in Hg1201, which, also belonging to the class of 90 K-$T_c$ cuprate material, is shown to display $T^2$-dependent resistivity as well in the pseudogap phase.\cite{Barisic2013} In Fig. \ref{fig-Hg120-RT-underdoped} we have reproduced the $\rho(T)$ measured by Yamamoto \textit{et al.} on Hg1201 polycrystals in the underdoped regime (from Ref. \onlinecite{Yamamoto2000}). From the temperature derivative d$\rho$/d$T$ plotted in the lower panels, we may readily perceive that the doping evolution of $\rho(T)$ in Hg1201 exhibits qualitatively similar features to that in YBCO thin films and single crystals (Figs. \ref{fig-summary-thinfilm} and \ref{fig-summary-single-crystal}), with the temperature range over which the normal-state resistivity follows a $T^2$ dependence (that is, the interval between $T_2$ and $T_f$) shrinks with increasing doping and becomes absent when $p$ reaches about 0.13. Intriguingly, it is shown by zero intercept of the linear d$\rho$/d$T$ at zero temperature (blue dashed lines in lower panels of Fig. \ref{fig-Hg120-RT-underdoped}) that the $\rho(T)$ exhibits a pure $T^2$ dependence in Hg1201 polycrystals for $0.05\leq p\leq0.11$, with no indication of either a low-$T$ upturn at low dopings or the presence of a $T$-linear term at high dopings in $\rho(T)$. This is apparently different from the situation in YBCO as discussed above. In fact, this phenomenon of unmixed $T^2$ dependence in underdoped Hg1201 has been noticed and confirmed in the study of $\rho_{ab}(T)$ on single crystal samples. \cite{Barisic2013} We note here that this subtle discrepancy in the doping evolution of $\rho_{ab}(T)$ between YBCO and Hg1201, whose origin may deserve future explorations, has implications on the transport properties of the two compounds. In particular, it could help further clarify a specific issue concerning the magnetotransport property of YBCO, that is, whether the normal-state transverse (orbital) magnetoresistance $\Delta\rho/\rho$ violates the Kohler's rule $\Delta\rho/\rho=F(H/\rho)$ ($F$ representing a scaling function) in the pseudogap phase. \cite{Harris1995,Chan2014} In a recent study, it has been shown that, in underdoped Hg1201 (at $p\simeq0.095$ and 0.11), the orbital magnetoresistance follows an $H^2T^{-4}$ dependence and hence obeys Kohler's rule in the temperature range over which the $\rho_{ab}(T)$ exhibits a $T^2$ dependence. \cite{Chan2014} It has been further suggested \cite{Chan2014} that the Kohler's rule is also obeyed in underdoped YBCO by noticing the similar presence of $T^2$-dependent resistivity in the pseudogap phase and the well-documented result that the orbital magnetoresistance (or more precisely, the $a$-axis magnetoconductivity) in YBCO follows an $H^2(aT^2+b)^{-2}$ dependence across the superconducting dome [$a$ and $b$ are coefficients with $b=0$ for oxygen content $y\geq6.60$ ($p\geq0.096$)]. \cite{Ando2002b} Here, a complement to this suggestion may be made. Given the established $H^2(aT^2+b)^{-2}$ behavior of the orbital magnetoresistance, the satisfaction of Kohler's rule in underdoped YBCO would require that the $\rho_{ab}(T)$ displayed a pure $T^2$ dependence as described by Eq. (\ref{eq-doping-regime-2}) in the corresponding temperature range. For $0.08\lesssim p\lesssim0.10$, this requirement is fulfilled for $T_f<T<T_2$, implying that the Kohler's rule is valid in this temperature range, as the above suggestion. \cite{Chan2014} It also applies for some of the single crystals of $0.10<p\lesssim0.11$ in which the $\rho_{ab}(T)$ still follows a pure $T^2$ dependence [Figs. \ref{fig-summary-single-crystal-B}(a) and \ref{fig-Rho-in-field-comparison}(b)]. For many other single crystals of $0.10<p\lesssim0.11$ and for $0.11<p<0.13$, however, owing to the presence of the $T$-linear term in $\rho_{ab}(T)$, the above requirement has not actually been met, indicating that the Kohler's rule should be violated in these samples despite the involvement of a $T^2$-dependent component in $\rho_{ab}(T)$. Indeed, we found that, if we used the $\rho_{a}(T)$ [Fig. \ref{fig-summary-single-crystal}(g)] and the magnetoresistance data for $y=6.70$ ($p=0.104$) single crystal as reported in Ref. \onlinecite{Ando2002b} to construct a Kohler plot, it would be seen that the Kohler's rule is not obeyed, including in the temperature range between $T_f$ and $T_2$. Hence, in underdoped YBCO, the satisfaction of Kohler's rule is limited to be in a narrower range of doping than that for the presence of a $T^2$-dependent component in $\rho_{ab}(T)$, and shows sample-dependent feature at certain dopings. We note that this seems to be a little bit more complicated than the situation in underdoped Hg1201 where, as mentioned earlier, the $\rho_{ab}(T)$ shows a pure $T^2$ dependence in temperatures between $T_f$ and $T_2$ for dopings $0.05\leq p\leq0.11$, \cite{Yamamoto2000,Barisic2013} thereby implying that the Kohler's rule should hold over the same doping range provided the $H^2(aT^2+b)^{-2}$ dependence of the magnetoresistance as demonstrated already at two dopings \cite{Chan2014} keeps in this range of doping.

The presence of a $T$-linear term coexisting with the $T$-quadratic in $\rho_{ab}(T)$ may remind us of the situation in overdoped high-$T_c$ cuprates. In Tl$_2$Ba$_2$CuO$_{6+\delta}$ (Tl2201), the normal-state $\rho_{ab}(T)$ at low $T$ accessed by employing magnetic fields was ever shown to contain a strong $T$-linear term aside from a $T^2$ term in samples of $T_c\simeq15$~K ($p\simeq0.26$), indicating a non-Fermi-liquid-like feature of the resistivity. \cite{Mackenzie1996,Proust2002} In recent years, such unusual temperature dependence of $\rho_{ab}(T)$ has been suggested to persist across the overdoped regime of this material, \cite{Hussey2013} and to arise possibly from the concurrence of two distinct scattering processes on the Fermi surface. \cite{Abdel-Jawad2006,Hussey2008} In overdoped LSCO, it has also been found that the low-$T$ normal-state $\rho_{ab}(T)$ extracted from high magnetic-field measurements could be expressed as a sum of $T$-linear and $T^2$-dependent components over a wide range of doping. \cite{Cooper2009,Hussey2011} These very similar phenomena seem to suggest an extended $T$-linearity of $\rho_{ab}(T)$ in overdoped high-$T_c$ cuprates, whose origin is currently unclear and scenarios have been brought forward. \cite{Cooper2009,Hussey2011,Hussey2013} In simply connecting these results with the observation of $\rho_{ab}(T)$ in YBCO for $0.10<p<0.13$, one may speculate that the coexistence of both $T$-linear and $T^2$ terms in low-$T$ normal-state $\rho_{ab}(T)$ may not be restricted in the overdoped regime, but also shows in the underdoped regime. This speculation seems to be backed by also noting that in overdoped LSCO it has been found that the coefficient $\alpha_1$ decreases while the $\alpha_2$ increases as doping lowers from about 0.19 towards 0.16, \cite{Cooper2009,Hussey2011} exhibiting the same trend of the opposite doping dependences of $\alpha_1$ and $\alpha_2$ in underdoped YBCO as shown in Fig. \ref{fig-summary-fit-coefficients} and mentioned earlier.

Owing to the onset of superconductivity, the shown behavior of the normal-state $\rho_{ab}(T)$ in YBCO below $T_2$ is confined in temperatures above $T_f$, and hence to actually check the above speculation, one needs to remove the superconductivity, for instance, by using high magnetic-fields, and explore the $\rho_{ab}(T)$ at the low-$T$ regime, as that performed in the overdoped cuprates. \cite{Mackenzie1996,Proust2002,Cooper2009} In other words, one needs to check out whether the $T$ dependence of $\rho_{ab}(T)$ between $T_2$ and $T_f$ persists down to the zero-temperature limit in underdoped YBCO, particularly for the doping part of $p>0.10$. To investigate this issue, we note that a closer look at the above fitting of $\rho_{ab}(T)$ between $T_2$ and $T_f$ may already provide an important clue. It is found that, for both YBCO thin films and single crystals, the yielded fitting coefficient $\rho_0$ [Fig. \ref{fig-summary-fit-coefficients}(a)] begins to show negative values when $p$ is above 0.10. This can be readily perceived from the upper panels of Figs. \ref{fig-summary-thinfilm}(f), \ref{fig-summary-single-crystal}(g), \ref{fig-summary-single-crystal}(h), and \ref{fig-summary-single-crystal-B}(b), or from the upper panels of Figs. \ref{fig-summary-single-crystal-B}(a) and \ref{fig-Rho-in-field-comparison}(b), in which the red lines representing the fits to Eq. (\ref{eq-doping-regime-3}) or (\ref{eq-doping-regime-2}), respectively, all show positive intercept on the horizontal $T$ axis implying that $\rho_0<0$. As in experiment one would expect a non-negative $\rho_0$ for $\rho_{ab}(T)$ at $T\rightarrow0$, this result suggests that the exact temperature dependence dictated by the fitting of $\rho_{ab}(T)$ between $T_2$ and $T_f$ would not extend down to the zero-temperature limit for $\rho_{ab}(T)$ at $p>0.10$. In the low-$T$ regime, the $\rho_{ab}(T)$ may follow a temperature dependence still described by Eq. (\ref{eq-doping-regime-3}) or (\ref{eq-doping-regime-2}) as that between $T_2$ and $T_f$ but the values of the coefficients should be changed; alternatively, the $\rho_{ab}(T)$ may transform to a temperature dependence different from the form of Eq. (\ref{eq-doping-regime-3}) or (\ref{eq-doping-regime-2}). This indicates the presence of crossover in the temperature evolution of $\rho_{ab}(T)$ at low $T$ for $p>0.10$ and, in particular, that the coexistence of $T$-linear and $T^2$ components in $\rho_{ab}(T)$ shown in many samples at $T_f<T<T_2$ could hold only at such intermediate temperatures. Interestingly, it is noted that the indication of crossover in $\rho_{ab}(T)$ at low $T$ also appears for underdoped Hg1201. As shown in the upper panel of Fig. \ref{fig-Hg120-RT-underdoped}(b), the fits of the resistivity above $T_f$ to Eq. \ref{eq-doping-regime-2} represented by blue dashed lines also have positive intercept on the $T$ axis for samples of $p=0.09$ and 0.11, giving $\rho_0<0$. Similar phenomenon is also shown in the $\rho_{ab}(T)$ of Hg1201 single crystals at $p=0.075$ and 0.11. \cite{Barisic2013} Such findings imply that the $\rho_{ab}(T)$ at these dopings of Hg1201 would undergo a crossover in the normal state exposed below $T_c$ and, similar to the case shown in Figs. \ref{fig-summary-single-crystal-B}(a) and \ref{fig-Rho-in-field-comparison}(b) for YBCO, if the $\rho_{ab}(T)$ continued to evolve with a $T^2$ dependence at the lowest $T$, the coefficient $\alpha_2$, that is, the strength of the $T^2$-dependence, should differ from the value exhibited at temperatures between $T_2$ and $T_f$.

\begin{figure}
  \includegraphics[scale=0.3]{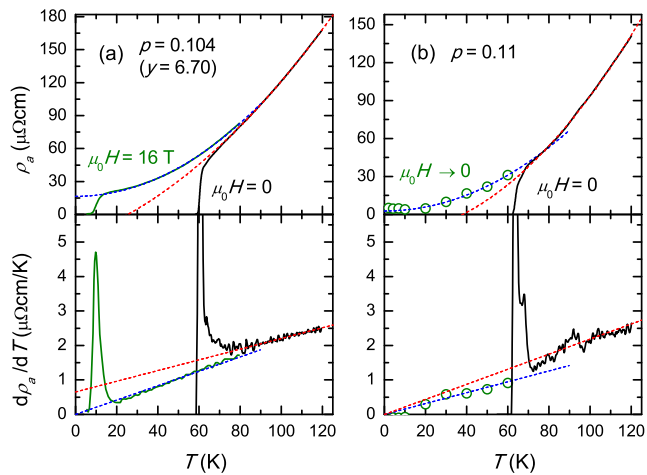}
  \caption{\label{fig-Rho-in-field-comparison}(Color online) Comparison of $\rho_{a}(T)$ of YBa$_2$Cu$_3$O$_y$ single crystals below 120 K in zero field and in finite field at two close dopings, $p=0.104$ ($y=6.70$) [(a), from Ref. \onlinecite{Ando2002b}] and 0.11 [(b), from Ref. \onlinecite{LeBoeuf2011}]. In (a), the $\rho_{a}(T)$ in zero field, as shown in Fig. \ref{fig-summary-single-crystal}(g), is fitted to Eq. (\ref{eq-doping-regime-3}) with $\rho_{0}<0$ for $T$ above $T_f$ (red dashed lines). While for the $\rho_{a}(T)$ in $\mu_0H=16$ T, it is shown by blue dashed lines that the data in the exposed normal state can be described by Eq. (\ref{eq-doping-regime-2}) with $\rho_{0}>0$, that is, by a pure $T^2$ dependence. In (b), the zero-field $\rho_{a}(T)$ is fitted to Eq. (\ref{eq-doping-regime-2}) as the approximate linear d$\rho_{a}$/d$T$ above $T_f$ shows zero intercept at $T=0$ K, giving $\rho_{0}<0$ (red dashed lines), and the $\rho_{a}(T)$ in $H\rightarrow0$ (obtained by measuring the normal-state magnetoresistance up to $\mu_0H=55$ T and extrapolating back to zero field)\cite{LeBoeuf2011} at lower temperatures could also be fitted to Eq. (\ref{eq-doping-regime-2}) but with $\rho_{0}>0$ and a different value of $\alpha_{2}$ (blue dashed lines).}
\end{figure}

In YBCO, we note that there are reports on the measurement of $\rho_{a}(T)$ in magnetic field below $T_c$ at dopings around 0.11, \cite{Ando2002b,LeBoeuf2011} which provide us a glimpse of the crossover in $\rho_{a}(T)$ at low $T$ as indicated from the fitting of the data between $T_2$ and $T_f$. In Fig. \ref{fig-Rho-in-field-comparison}(a), the zero-field $\rho_{a}(T)$ for $y=6.70$ ($p=0.104$), as shown in Fig. \ref{fig-summary-single-crystal}(g), has been compared with the $\rho_{a}(T)$ in $\mu_0H=16$ T measured on the same single crystal. \cite{Ando2002b} From the plot of d$\rho_{a}$/d$T$, it can be inferred that the moderate magnetic field has suppressed the superconductivity down to about 20 K and, moreover, the measured $\rho_{a}(T)$ between 20 K and 80 K follows a pure $T^2$ dependence as the linear d$\rho_{a}$/d$T$ shows zero intercept at $T=0$ K as indicated by the blue dashed line. The fit of $\rho_{a}(T)$ in this temperature range to Eq. (\ref{eq-doping-regime-2}) gives $\rho_0>0$ as shown by the blue dashed line in the upper panel, suggesting that the manifested $T^2$ dependence may persist down to $T=0$ K. Although this measured $\rho_{a}(T)$ in field of 16 T should contain a small contribution of the normal-state magnetoresistance which has not yet been removed, \cite{Cooper2009} its $T$ dependence indicates that the normal-state $\rho_{a}(T)$ of this sample is likely to cross over from the form of Eq. (\ref{eq-doping-regime-3}) at $T_f<T<T_2$ to the form of Eq. (\ref{eq-doping-regime-2}) at the lowest $T$ and $H\rightarrow0$.
In Fig. \ref{fig-Rho-in-field-comparison}(b), besides the zero-field $\rho_{a}(T)$, the magnetoresistance-free ($H\rightarrow0$) normal-state $\rho_{a}(T)$ below $T_c$ is also shown for a single crystal of $p=0.11$, which has been extracted by measuring the $\rho_{a}(H)$ up to $\mu_0H=55$ T at various low temperatures down to 2 K and then extrapolating the normal-state part back to zero field. \cite{LeBoeuf2011} The blue dashed lines indicate that it follows approximately a pure $T^2$ dependence with a very small $\rho_0$. For this sample, it is shown that the zero-field $\rho_{a}(T)$ above $T_c$ also exhibits roughly a pure $T^2$ dependence at temperatures between 120 K and $T_f\simeq72$ K, although the fit to Eq. (\ref{eq-doping-regime-2}) yields $\rho_0<0$ and a different $\alpha_2$, as indicated by the red dashed lines. Hence, for this crystal the crossover in the normal-state $\rho_{a}(T)$ shows as a change in the strength of the $T^2$-dependence as temperature drops below $T_c$. From both cases shown in Fig. \ref{fig-Rho-in-field-comparison}, we see that, at lowest $T$, a pure $T^2$-dependent $\rho_{a}(T)$ seems to be consistently pointed to for YBCO at $p\sim0.11$. As doping further proceeds, whether this Fermi liquid-like behavior of $\rho_{a}(T)$ persists and if so, to which doping it extends would be very appealing to explore in more systematic high-field experiments at low $T$ in the underdopd YBCO. This would also allow for a more direct comparison to the results shown in the overdoped high-$T_c$ cuprates. \cite{Mackenzie1996,Proust2002,Cooper2009} In this respect, we note that recently the $\rho_{a}(T)$ has been reported to exhibit a $T^2$ dependence below $T_c$ in underdoped YBa$_2$Cu$_4$O$_8$ ($T_c\sim80$ K) based on $\rho_{a}(H)$ measurements in magnetic fields up to 60 T at temperatures down to 1.5 K. \cite{Proust2016}

\begin{figure}
  \includegraphics[scale=0.28]{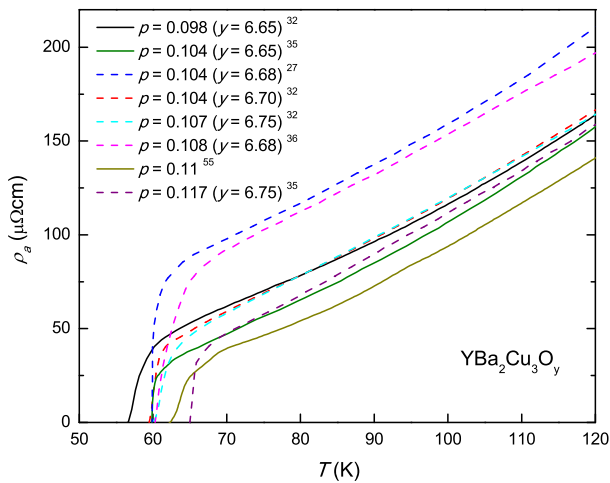}
  \caption{\label{fig-summary-single-crystal-C}(Color online) Compilation of $\rho_{a}(T)$ of YBa$_2$Cu$_3$O$_y$ single crystals below 120 K as typically reported at $0.10\lesssim p\lesssim0.12$ (from Refs. \onlinecite{Ito1993,Takenaka1994,Segawa2001,Lee2005,LeBoeuf2011}). The dashed lines are the $\rho_{a}(T)$ showing the presence of a $T$-linear component at $T_f<T<T_2$, some of which illustrated in Figs. \ref{fig-summary-single-crystal}(g), \ref{fig-summary-single-crystal}(h), and \ref{fig-summary-single-crystal-B}(b). The solid lines are the $\rho_{a}(T)$ showing a pure $T^2$ dependence at $T_f<T<T_2$, as illustrated in Figs. \ref{fig-summary-single-crystal}(f), \ref{fig-summary-single-crystal-B}(a), and \ref{fig-Rho-in-field-comparison}(b).}
\end{figure}

Concerning the $T$-linear component, another intriguing issue is the sample dependence of its presence or absence in $\rho_{ab}(T)$ for $T_f<T<T_2$ at $0.10<p\lesssim0.11$, as demonstrated above. A natural question is posed as why it shows in some single crystals but not in other crystals having the same or similar level of doping ($T_c$). To find a clue, in Fig. \ref{fig-summary-single-crystal-C} we have plotted the $\rho_{a}(T)$ of these crystals together. It is interesting to note that, compared with the crystals showing the presence of $T$-linear component in $\rho_{a}(T)$ at $p>0.10$ (dashed lines), the two crystals showing a pure $T^2$ dependence at the same or close dopings ($p=0.104$ and 0.11, solid lines) have lower values of $\rho_{a}(T)$ in temperatures between $T_c$ and 120 K. This seems to indicate a lower disorder level in these two crystals and that in other crystals the presence of a $T$-linear component in $\rho_{a}(T)$ might have a connection with the slightly higher level of disorder. In other words, the disorder in YBCO single crystals might possibly have a novel effect in promoting the appearance of $T$-linear resistivity and in some crystals showing no $T$-linear component in $\rho_{a}(T)$ the level of disorder might just fall below a certain threshold. It is noted that, previously, in discussing the origin of the extended $T$-linearity of $\rho_{ab}(T)$ in overdoped LSCO, the possible role of disorder has also been considered. \cite{Cooper2009} To explore further such a conjecture, more experiments on YBCO crystals of varying disorder would be needed to clarify the issues such as what the nature of the disorder is and why it could correlate with the $T$-linear dependence of $\rho_{a}(T)$. Here, one may be reminded that, at $0.08\lesssim p\lesssim0.10$, the $\rho_{ab}(T)$ also varies in magnitude in different YBCO single crystals and thin films of similar dopings, and yet retains a pure $T^2$ dependence in these samples, seemingly showing insensitive to moderate changes of disorder level and hence in contrast with the situation shown in Fig. \ref{fig-summary-single-crystal-C} for $p>0.10$.

Having the doping variation of the temperature $T_2$, combined with that of the temperatures $T_f$ and $T^\ast$, it may be appropriate to point out in particular that the experimentally determined phase diagram shown in Fig. \ref{fig-Phasediagram} for YBCO looks similar to that recently proposed theoretically by Lee for high-$T_c$ cuprates. \cite{Lee2014} In the theoretical phase diagram [Fig. 7(a) in Ref. \onlinecite{Lee2014}], at temperatures immediately above the superconducting dome, there is a regime for fluctuating $d$-wave superconductivity illustrated to be tied to $T_c$, similar to the narrow regime of superconducting fluctuations shown in Fig. \ref{fig-Phasediagram} defined by the $T_f$. The regime at higher temperatures, namely the pseudogap phase, is theoretically characterized as a novel pair-density-wave state with the onset of pairing amplitude at temperature $T_{\mathrm{PDW}-\mathrm{MF}}$, \cite{Lee2014} which could be viewed as the pseudogap temperature $T^\ast$ in experiment. Apparently, whether the pseudogap phase could be identified as a pair-density-wave state could not be known from the present study. Below $T_{\mathrm{PDW}-\mathrm{MF}}$, there resides another characteristic temperature inside the pseudogap phase, which, decreasing with doping in parallel with $T_{\mathrm{PDW}-\mathrm{MF}}$, is predicted as the onset of an induced charge-density-wave (CDW) order and hence labelled as $T_{\mathrm{CDW}}$. \cite{Lee2014} Intriguingly, the position of $T_{\mathrm{CDW}}$ in the phase diagram and its doping dependence, although shown schematically, look rather like that of the temperature $T_2$ determined from the resistivity, suggesting a possible correspondence between the two temperatures. Although it is highly speculative, the above suggestion of the $T_2$ corresponding to the onset of a CDW order may carry importance in illuminating the origin of the emergence of $T^2$-dependent resistivity in the pseudogap phase. In this respect, we note, interestingly, that, at certain doping intervals the temperature $T_2$ obtained in the present study is indeed close to the temperature $T_{\mathrm{CDW}}$ that determined in recent x-ray diffraction/scattering experiments for YBCO, \cite{Blanco-Canosa2014,Hucker2014} as will be discussed in the next subsection.

\subsection{\label{subsec:Res-Hall-Seebeck}Correlation between resistivity, Hall coefficient, and the Seebeck coefficient}

\begin{figure*}
  \includegraphics[scale=0.62]{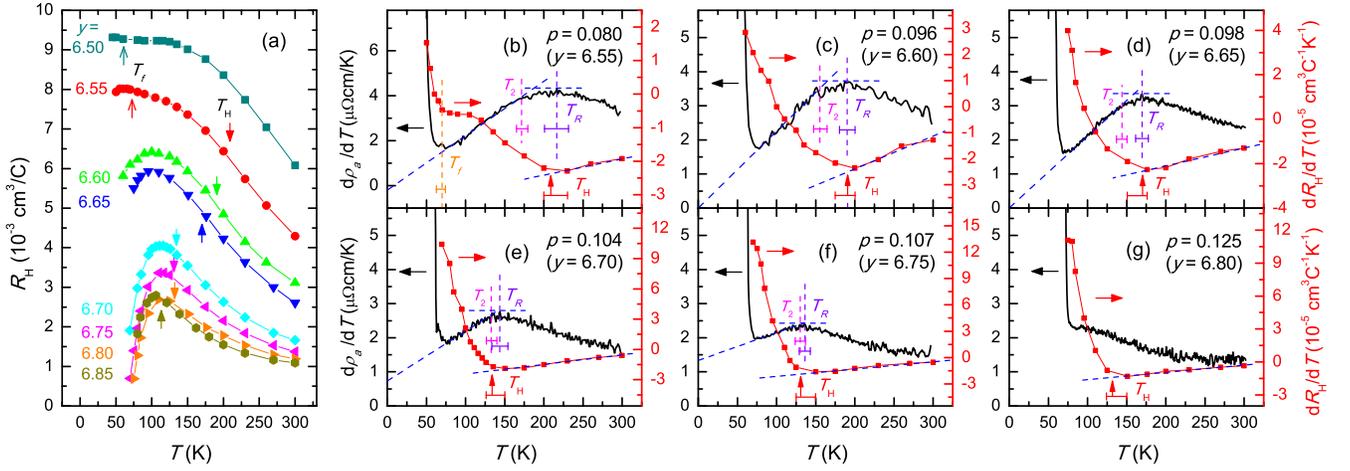}
  \caption{\label{fig-summary-Res-and-Hall}(Color online) (a) Temperature dependence of the in-plane Hall coefficient $R_\mathrm{H}$ of YBa$_2$Cu$_3$O$_y$ single crystals with $y$ values as indicated (from Ref. \onlinecite{Segawa2004}). (b)-(g) Doping evolution of the temperature derivative d$R_\mathrm{H}$/d$T$ of the $R_\mathrm{H}(T)$ shown in (a), in comparison with that of the d$\rho_{a}$/d$T$ as depicted in Fig. \ref{fig-summary-single-crystal}. In each panel of (b)-(g), the d$R_\mathrm{H}$/d$T$ is shown to decrease approximately linearly as $T$ decreases from room temperature (as illustrated by a blue diagonal dashed line), until arriving at a temperature $T_\mathrm{H}$ (denoted by a red vertical arrow with horizontal error bar), below which the d$R_\mathrm{H}$/d$T$ increases instead. The $T_\mathrm{H}$ therefore indicates where the $R_\mathrm{H}(T)$ changes from upward to downward curvature as temperature is reduced. Intriguingly, it is shown in (b)-(f) that the $T_\mathrm{H}$ coincides well with the temperature $T_R$ (indicated by a violet vertical dashed line) at which the d$\rho_{a}$/d$T$ reaches a local maximum (indicated by a blue horizontal dashed line), i.e., an inflection point in the $\rho_{a}(T)$ curve where it goes from downward to upward curvature as temperature is reduced. From (b) to (f), it is shown that the temperature $T_2$ defined in Fig. \ref{fig-summary-single-crystal} gradually merges into the temperature $T_R$ as doping increases. When $p$ reaches around 0.13 or above, the $\rho_{a}(T)$ shows no inflection point between $T^\ast$ and $T_c$, and hence a definition of $T_R$ has not been made and the correspondence between $T_\mathrm{H}$ and $T_R$ is lost, as shown in (g). In (b), the d$R_\mathrm{H}$/d$T$ curve for $y=6.55$ ($p=0.080$) shows another knee point at a lower temperature, which is seen to coincide with the temperature $T_f$ as defined in Fig. \ref{fig-summary-single-crystal} from the d$\rho_{a}$/d$T$ (as indicated by an orange vertical dashed line). Similar phenomenon is shown in the crystal of $y=6.50$ ($p=0.074$), as indicated in (a) by the arrows labelled as $T_f$. Other arrows in (a) at higher temperatures indicate positions of the temperature $T_\mathrm{H}$.}
\end{figure*}

To search for the origin of the change of $\rho_{ab}(T)$ into a prominent $T^2$ dependence below $T_2$, it is desirable to see first whether there are also changes appeared at temperatures close to $T_2$ in other electrical transport properties of YBCO. For this, in Fig. \ref{fig-summary-Res-and-Hall}(a) we have reproduced the in-plane Hall coefficient $R_\mathrm{H}(T)$ in a wide doping range of YBCO (from Ref. \onlinecite{Segawa2004}) and investigated its temperature dependence at each doping by plotting the derivative d$R_\mathrm{H}$/d$T$ in Figs. \ref{fig-summary-Res-and-Hall}(b) to \ref{fig-summary-Res-and-Hall}(g). Because these systematic $R_\mathrm{H}(T)$ data were obtained by Ando \textit{et al.} on the same crystals as that used to obtain the $a$-axis resistivity $\rho_a(T)$ as displayed in Fig. \ref{fig-summary-single-crystal}, in Figs. \ref{fig-summary-Res-and-Hall}(b) to \ref{fig-summary-Res-and-Hall}(g) we have included the corresponding d$\rho_{a}$/d$T$ curve as well to make a direct comparison between the d$R_\mathrm{H}$/d$T$ and d$\rho_{a}$/d$T$. It is shown that, with doping varying from 0.080 to 0.139 ($y=6.85$), the evolution of d$R_\mathrm{H}$/d$T$ with temperature remains similar, that is, the d$R_\mathrm{H}$/d$T$ first decreases approximately linearly with temperature decreasing from 300 K and then increases relatively quickly with temperature decreasing further, exhibiting a minimum at a doping-dependent temperature labelled as $T_\mathrm{H}$ in each panel of Figs. \ref{fig-summary-Res-and-Hall}(b) to \ref{fig-summary-Res-and-Hall}(g). This defined $T_\mathrm{H}$ therefore denotes the point where the $R_\mathrm{H}(T)$ changes from upward to downward curvature as temperature is reduced. Remarkably, it is revealed in Figs. \ref{fig-summary-Res-and-Hall}(b) to \ref{fig-summary-Res-and-Hall}(f) that at these dopings the $T_\mathrm{H}$ coincides very well with the temperature at which the d$\rho_{a}$/d$T$ shows a local maximum, labelled as $T_R$ in the figures. As this $T_R$ denotes similarly the point where the $\rho_{a}(T)$ curve changes curvature, i.e., goes from downward to upward curvature as temperature is reduced, \cite{DRemark} it tells us that there is a good correspondence between the inflection points in $R_\mathrm{H}(T)$ and $\rho_{a}(T)$ curves for certain doping intervals of underdoped YBCO. We note that this correspondence could also be observed in the $R_\mathrm{H}(T)$ and $\rho_{ab}(T)$ data of epitaxial thin films \cite{Wuyts1993,Wuyts1996} (see Fig. \ref{fig-various-T-comparison}). As shown in Figs. \ref{fig-summary-Res-and-Hall}(b) to \ref{fig-summary-Res-and-Hall}(f), the temperature $T_R$ precedes the temperature $T_2$ as defined in Fig. \ref{fig-summary-single-crystal} and discussed above and the two temperatures become closer and closer and finally merge into each other as doping increases. Through the coincidence between $T_R$ and $T_\mathrm{H}$, this indicates that there should be a correlation between the appearance of downward curvature in $R_\mathrm{H}(T)$ below $T_\mathrm{H}$ and the presence of $T^2$ dependence in $\rho_{ab}(T)$ below $T_2$.

\begin{figure*}
  \includegraphics[scale=0.62]{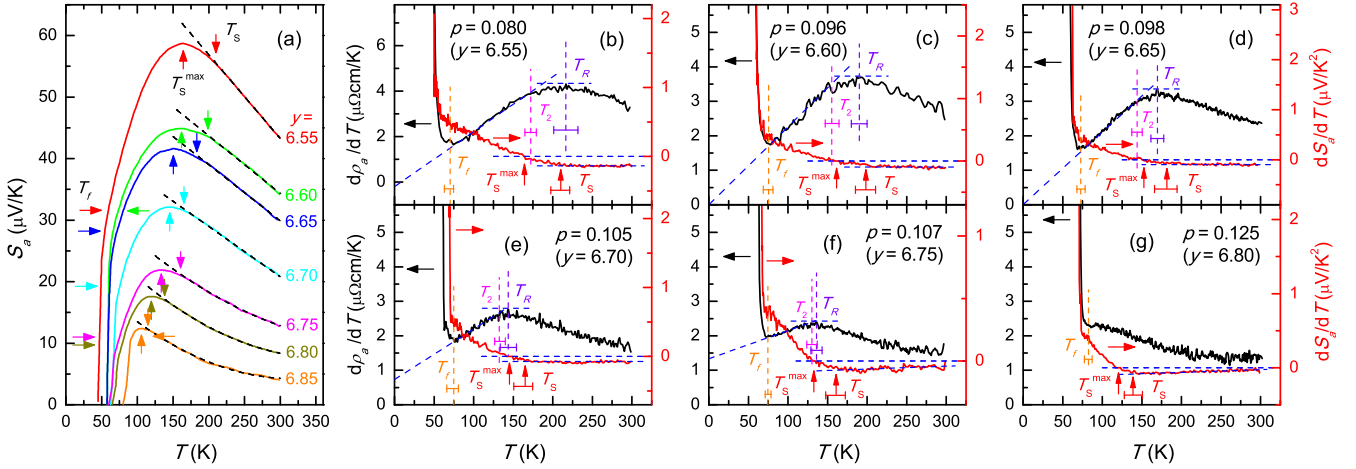}
  \caption{\label{fig-summary-Res-and-TEP}(Color online) (a) Temperature dependence of the $a$-axis Seebeck coefficient or, equivalently, the thermopower $S_a$ of YBa$_2$Cu$_3$O$_y$ single crystals with $y$ values as indicated (from Ref. \onlinecite{Segawa2003}). (b)-(g) Doping evolution of the temperature derivative d$S_a$/d$T$ of the $S_a(T)$ shown in (a), in comparison with that of the d$\rho_{a}$/d$T$ as depicted in Fig. \ref{fig-summary-single-crystal}. In each panel of (b)-(g), the d$S_a$/d$T$ is shown to be temperature-independent [(b)-(e)] or decrease approximately linearly [(f) and (g)] as $T$ decreases from 300 K (as illustrated by a horizontal or diagonal blue dashed line), until arriving at a temperature $T_\mathrm{S}$ (denoted by a red vertical arrow with horizontal error bar), below which the d$S_a$/d$T$ starts to increase. Thus the $T_\mathrm{S}$ [marked in (a) by down arrows] indicates where the $S_a(T)$ begins to deviate from its high-$T$ behavior as temperature is reduced [developing or changing to a downward curvature as illustrated in (a) by black dashed lines]. It is shown in (b)-(f) that the $T_\mathrm{S}$ is quite close to the temperature $T_R$ (indicated by a violet vertical dashed line), as defined in Fig. \ref{fig-summary-Res-and-Hall}, at which the d$\rho_{a}$/d$T$ reaches a local maximum (indicated by a blue horizontal dashed line). As $T$ decreases from $T_\mathrm{S}$, the d$S_a$/d$T$ increases to become zero at a temperature $T_\mathrm{S}^{\mathrm{max}}$ as denoted in each panel of (b)-(g), which corresponds to the $S_a(T)$ reaching a local maximum [marked in (a) by up arrows]. It is shown in (b)-(f) that the $T_\mathrm{S}^{\mathrm{max}}$ matches well with the temperature $T_2$ (indicated by a magenta vertical dashed line) as defined in Fig. \ref{fig-summary-single-crystal} for $\rho_{a}(T)$. In each panel of (b)-(g), it is also shown that, below $T_\mathrm{S}^{\mathrm{max}}$, the increase of d$S_a$/d$T$ with decreasing $T$ displays an abrupt change at a certain temperature, below which the increase of d$S_a$/d$T$ becomes much steeper. It is shown that this temperature, marking the point where the fall of $S_a(T)$ with decreasing $T$ suddenly becomes much faster [indicated in (a) by horizontal arrows], coincides well with the temperature $T_f$ as defined in Fig. \ref{fig-summary-single-crystal} from the d$\rho_{a}$/d$T$ curve (as indicated by an orange vertical dashed line).}
\end{figure*}

We have also examined another electrical transport property, i.e., the Seebeck coefficient or, equivalently, the thermopower $S$ of YBCO and compared its temperature dependence with that of the $\rho_{a}(T)$ in the underdoped regime, as demonstrated in Fig. \ref{fig-summary-Res-and-TEP}. Here, the $a$-axis $S_a(T)$ data from 300 K down to $T_c$, taken from Ref. \onlinecite{Segawa2003} and reproduced in Fig. \ref{fig-summary-Res-and-TEP}(a), were also measured systematically by Segawa and Ando on the YBCO single crystals that used for the $\rho_{a}(T)$ and $R_\mathrm{H}(T)$ measurements as shown in Figs. \ref{fig-summary-single-crystal} and \ref{fig-summary-Res-and-Hall}. It is seen from Fig. \ref{fig-summary-Res-and-TEP}(a) that, as temperature is reduced from 300 K, the $S_a(T)$ initially increases roughly linearly or with an upward curvature (as indicated by black dashed lines) and then starts to bend down and develops a downward curvature at a temperature $T_\mathrm{S}$ (marked by down arrows), corresponding to the d$S_a$/d$T$ being nearly temperature-independent or decreasing linearly at first and then starting to increase for $T$ below $T_\mathrm{S}$, as shown in Figs. \ref{fig-summary-Res-and-TEP}(b) to \ref{fig-summary-Res-and-TEP}(g). As temperature reduces further, the $S_a(T)$ forms a peak at a temperature $T_\mathrm{S}^{\mathrm{max}}$ [marked in Fig. \ref{fig-summary-Res-and-TEP}(a) by up arrows], with the increase of d$S_a$/d$T$ crossing the zero value as illustrated in Figs. \ref{fig-summary-Res-and-TEP}(b) to \ref{fig-summary-Res-and-TEP}(g). By comparing the d$S_a$/d$T$ with d$\rho_{a}$/d$T$, it is intriguing to see from Figs. \ref{fig-summary-Res-and-TEP}(b) to \ref{fig-summary-Res-and-TEP}(f) that at these dopings the temperature $T_\mathrm{S}$ is quite close to the temperature $T_R$ and the temperature $T_\mathrm{S}^{\mathrm{max}}$ coincides well with the temperature $T_2$, \cite{ERemark} suggesting that the bending down and the subsequent appearance of a maximum in $S_a(T)$ have an intimate relation to the appearance of the curvature change and then the $T^2$ dependence in $\rho_{a}(T)$. Together with the coincidence between $T_\mathrm{H}$ and $T_R$ as shown in Figs. \ref{fig-summary-Res-and-Hall}(b) to \ref{fig-summary-Res-and-Hall}(f), this establishes an agreement among $T_R$, $T_\mathrm{H}$, and $T_\mathrm{S}$, as summarized in Fig. \ref{fig-various-T-comparison}(a), which strongly indicates that it should be the same mechanism that leads to the inflection points or downward curvature appeared concurrently in $\rho_{a}(T)$, $R_\mathrm{H}(T)$, and $S_a(T)$ in underdoped YBCO. And it should also be this mechanism that is responsible for the emergence of $T^2$-dependent resistivity in the pseudogap phase of YBCO.

It is also shown from Figs. \ref{fig-summary-Res-and-TEP}(b) to \ref{fig-summary-Res-and-TEP}(g) that, as temperature is decreased further from $T_\mathrm{S}^{\mathrm{max}}$, at each doping the d$S_a$/d$T$ initially increases steadily, but then exhibits abruptly a much steeper increase, showing an apparent upturn at low $T$. The presence of this upturn in d$S_a$/d$T$ has actually revealed another feature of the $S_a(T)$ data shown in Fig. \ref{fig-summary-Res-and-TEP}(a), that is, the decrease of $S_a(T)$ becomes much faster as temperature reduces to be below a certain point, as indicated by horizontal arrows in Fig. \ref{fig-summary-Res-and-TEP}(a). By comparing with the d$\rho_{a}$/d$T$, it is remarkable to see in Figs. \ref{fig-summary-Res-and-TEP}(b) to \ref{fig-summary-Res-and-TEP}(g) that the onset of the upturn in d$S_a$/d$T$ coincides well with the onset of the upturn in d$\rho_{a}$/d$T$, which, defined as the temperature $T_f$, has been ascribed to the onset of superconducting fluctuations as discussed in Sec. \ref{subsec:SC-fluc}. This coincidence accordingly indicates that the onset of the upturn in d$S_a$/d$T$ should also be linked with the appearance of superconducting fluctuations. In other words, it should be the onset of fluctuating superconductivity that leads to the $S_a(T)$ starting to fall more rapidly as $T$ goes down to around $T_f$. This reveals the effect of the emergence of superconducting fluctuations on the behavior of $S_a(T)$ at low temperatures and suggests that it is also a convenient method to plot the d$S_a$/d$T$ curve and identify the onset of the upturn in it to determine the temperature range of superconducting fluctuations above $T_c$, similar to the case with d$\rho_{ab}$/d$T$. In view of this, in Fig. \ref{fig-Phasediagram} we have included the $T_f$ determined from the $S_a(T)$ shown in Fig. \ref{fig-summary-Res-and-TEP} and the in-plane $S_{ab}(T)$ data reported by Wang and Ong (Ref. \onlinecite{Wang2001}) in which the d$S_{ab}$/d$T$ also exhibits clear upturn above $T_c$. In fact, not only in $\rho_{ab}(T)$ and $S_{ab}(T)$, in the Hall coefficient $R_\mathrm{H}(T)$ the trace of superconducting fluctuations may also be spotted by examining the temperature derivative d$R_\mathrm{H}$/d$T$, although in experimentally determining $R_\mathrm{H}(T)$ close to $T_c$ one would usually have paid attention to using the linear-in-$H$ regime of the Hall resistivity to try to avoid the influence of superconducting fluctuations. For instance, it is seen in Fig. \ref{fig-summary-Res-and-Hall}(b) that, for sample of $y=6.55$ ($p=0.08$), the $R_\mathrm{H}(T)$ has been measured down to temperatures very close to $T_c$ and the d$R_\mathrm{H}$/d$T$ exhibits a knee structure in the low-$T$ regime showing an upturn below the knee point. Importantly, as shown in Fig. \ref{fig-summary-Res-and-Hall}(b), this upturn in d$R_\mathrm{H}$/d$T$ appears precisely at the temperature $T_f$ which marks the onset of the upturn in d$\rho_{a}$/d$T$. The very similar phenomenon has been found for $y=6.50$ ($p=0.074$) sample as well. These suggest that the presence of the upturn in d$R_\mathrm{H}$/d$T$ should also be a manifestation of the appearance of superconducting fluctuations. Through the above comparisons, we can see that, besides the correlation on the inflection point or curvature change at relatively high temperatures, the $\rho_{ab}(T)$, $S_{ab}(T)$, and $R_\mathrm{H}(T)$ in underdoped YBCO have also demonstrated another correlation at relatively low temperatures, that is, the development of an upturn in d$\rho_{ab}$/d$T$, d$S_{ab}$/d$T$, or d$R_\mathrm{H}$/d$T$ at nearly the same temperatures. And it is the onset of superconducting fluctuations that lies behind this lower-temperature correlation among the three transport properties. These findings further illustrate the usefulness of plotting the derivatives to unveil valuable features or connections of such physical quantities in high-$T_c$ cuprates.

\begin{figure*}
  \includegraphics[scale=0.6]{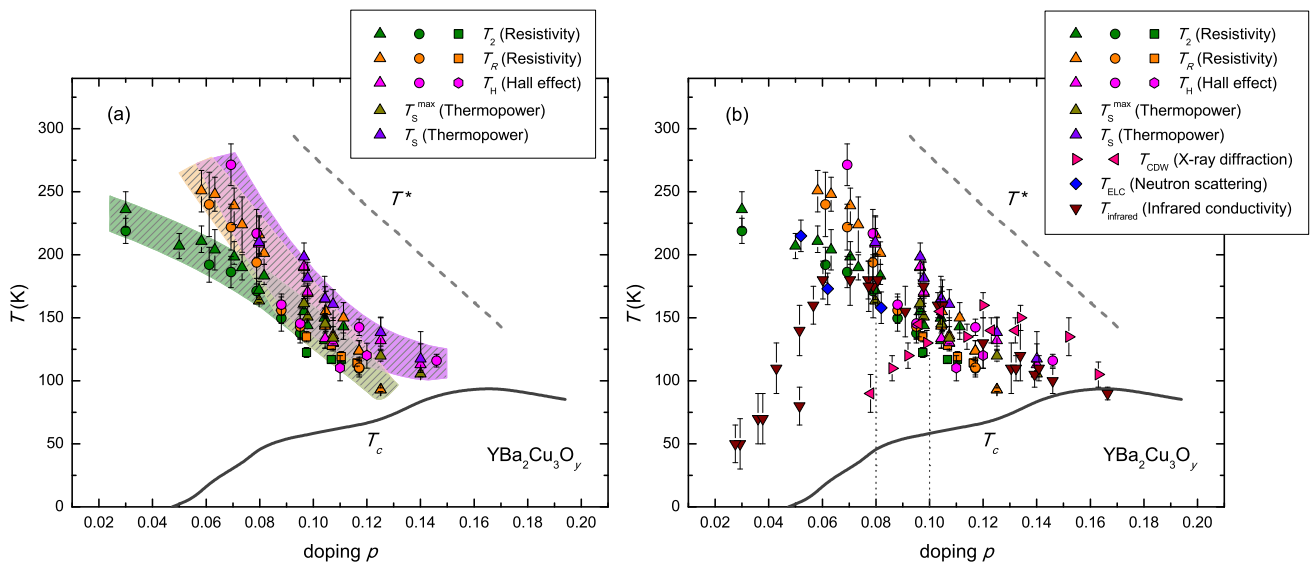}
  \caption{\label{fig-various-T-comparison}(Color online) (a) Comparing the characteristic temperatures $T_2$ and $T_R$ extracted from resistivity measurement (as shown in Fig. \ref{fig-Phasediagram}) with the $T_\mathrm{H}$ defined from the Hall effect $R_\mathrm{H}(T)$ experiment (as explained in Fig. \ref{fig-summary-Res-and-Hall}) and the $T_\mathrm{S}^{\mathrm{max}}$ and $T_\mathrm{S}$ from the thermopower $S_a(T)$ experiment (as explained in Fig. \ref{fig-summary-Res-and-TEP}). Error bars are explained in Figs. \ref{fig-Phasediagram}, \ref{fig-summary-Res-and-Hall}, and \ref{fig-summary-Res-and-TEP}, or quoted from Ref. \onlinecite{LeBoeuf2011}. For $T_\mathrm{H}$, the magenta up triangles and circles represent the results obtained from $R_\mathrm{H}(T)$ measurements on single crystals \cite{Segawa2004} (shown in Fig. \ref{fig-summary-Res-and-Hall}) and thin films \cite{Wuyts1996} respectively, and the magenta hexagons are the data taken from Ref. \onlinecite{LeBoeuf2011}. Shaded areas are guides to the eye. The solid line indicates the $T_c$ from Ref. \onlinecite{Liang2006} and the dashed line is a schematic plot of the pseudogap temperature $T^\ast$ following the data shown in Fig. \ref{fig-Phasediagram}.   It is seen that the temperature $T_2$ closely tracks the $T_R$ as doping decreases down to about 0.08 and then deviates from the trajectory of the latter as doping further decreases, corresponding to the observation that the maximum in d$\rho_{ab}$/d$T$ at $T_R$ becomes shallow and broad with doping reducing into the deeply underdoped regime, as displayed in Figs. \ref{fig-summary-single-crystal} and \ref{fig-summary-Res-and-Hall}. The temperatures $T_R$, $T_\mathrm{H}$, and $T_\mathrm{S}$ coincide with each other for $p\lesssim0.12$, and with doping further increasing, the coincidence between $T_\mathrm{H}$ and $T_\mathrm{S}$ remains while the definition of $T_R$ becomes unapplicable, as demonstrated in Figs. \ref{fig-summary-Res-and-Hall}(g) and \ref{fig-summary-Res-and-TEP}(g). (b) The same as in (a), plus the plot of several characteristic temperatures inside the pseudogap phase of YBCO identified from other kinds of experiments. The $T_{\mathrm{CDW}}$ is the temperature below which charge-density-wave order is detected in x-ray diffraction measurements (right triangles from Ref. \onlinecite{Blanco-Canosa2014} and left triangles from Ref. \onlinecite{Hucker2014}). The temperature $T_{\mathrm{ELC}}$ represents the transition to an electronic-liquid-crystal state as reported by neutron scattering (diamonds, from Ref. \onlinecite{Haug2010}). The temperature $T_{\mathrm{infrared}}$ is defined as the onset of anomalous softening or broadening of certain infrared-active phonons as seen in infrared $c$-axis conductivity (down triangles, from Ref. \onlinecite{Dubroka2011}). The associated error bars are quoted from the references cited. \cite{Blanco-Canosa2014,Hucker2014,Haug2010,Dubroka2011} Two vertical dotted lines mark the doping points $p=0.08$ and 0.10 respectively.}
\end{figure*}

To elucidate what lies behind the higher-temperature correlation among $\rho_{ab}(T)$, $R_\mathrm{H}(T)$, and $S_{ab}(T)$, namely the simultaneous appearance of inflection points or downward curvature in them, it is instructive to compare the characteristic temperatures shown in them with possible similar temperature scales identified from other kinds of experiments. In Fig. \ref{fig-various-T-comparison}(b), besides the temperatures $T_2$ and $T_R$ from $\rho_{ab}(T)$, the $T_\mathrm{H}$ from $R_\mathrm{H}(T)$, and the $T_\mathrm{S}^{\mathrm{max}}$ and $T_\mathrm{S}$ from $S_{ab}(T)$ as shown in Fig. \ref{fig-various-T-comparison}(a), we have also depicted the doping evolution of some characteristic temperatures including $T_{\mathrm{CDW}}$, $T_{\mathrm{ELC}}$, and $T_{\mathrm{infrared}}$ determined previously from several other types of measurements in YBCO. Here, the $T_{\mathrm{CDW}}$ represents the onset temperature for CDW order as obtained in recent soft or hard x-ray diffraction experiments, \cite{Blanco-Canosa2014,Hucker2014} the $T_{\mathrm{ELC}}$ denotes the temperature below which an electronic-liquid-crystal state as seen in neutron scattering experiment appears, \cite{Haug2010} and the $T_{\mathrm{infrared}}$, as mentioned earlier, is defined from infrared $c$-axis conductivity as the onset temperature for anomalous softening or broadening of certain infrared-active phonons in YBCO. \cite{Dubroka2011} It is shown in Fig. \ref{fig-various-T-comparison}(b) that there is significant overlap among these temperatures over certain doping intervals. The temperature $T_{\mathrm{ELC}}$, covering $0.05\lesssim p\lesssim0.08$, follows the line of $T_2$ and $T_\mathrm{S}^{\mathrm{max}}$. The $T_{\mathrm{CDW}}$, which spans over $0.08\lesssim p\lesssim0.16$, is shown to increase in parallel with other temperatures like the $T_\mathrm{H}$ and $T_\mathrm{S}$ as doping drops from about 0.16 to about $0.12-0.10$, and then, in opposite to other temperatures, decrease steadily as doping further reduces towards about 0.08.  The temperature $T_{\mathrm{infrared}}$, defined across the range of $0.03\lesssim p\lesssim0.17$, also shows good agreement with the temperatures like $T_\mathrm{H}$ and $T_\mathrm{S}$ when $p$ is above $\sim0.08-0.06$. For $p$ less than this value, the $T_{\mathrm{infrared}}$ changes to decrease roughly linearly with decreasing doping, contrary to the behavior of other temperatures like $T_2$ and $T_{\mathrm{ELC}}$.

The above observation of significant overlap among various characteristic temperatures hints at their possible internal connections. Specifically, it indicates that the presence of inflection points or curvature change in the three transport quantities might be related to the onset of modulations in charge or spin correlations in the pseudogap phase of YBCO. In recent studies,\cite{LeBoeuf2007,LeBoeuf2011} by using high magnetic fields to suppress superconductivity, a drop and sign change of $R_\mathrm{H}$ into negative values, following its drop at temperatures above $T_c$ as displayed in Fig. \ref{fig-summary-Res-and-Hall}(a), has been uncovered in the exposed normal state at low temperatures for YBCO at intermediate dopings, suggestive of a Fermi-surface reconstruction with the formation of electron pocket in the underdoped YBCO as implied by quantum oscillation experiments. \cite{Sebastian2015} According to this, the $T_\mathrm{H}$, marking the beginning of downward curvature in $R_\mathrm{H}(T)$, was ever viewed as the onset temperature for such Fermi-surface reconstruction. \cite{LeBoeuf2011} Upon the discovery of CDW modulations by x-ray diffractions, \cite{Ghiringhelli2012,Chang2012} it was immediately noticed that its onset temperature $T_{\mathrm{CDW}}$ at $p=0.12$ corresponds approximately with $T_\mathrm{H}$, \cite{Chang2012} in support of the idea that the suggested Fermi-surface reconstruction could be caused by the CDW modulations. \cite{Sebastian2015} It is shown in Fig. \ref{fig-various-T-comparison}(b) that the approximate correspondence between $T_{\mathrm{CDW}}$ and $T_\mathrm{H}$ actually extends for dopings in the range of $0.10\lesssim p\lesssim0.16$, which seems to back further the above idea. Following such a line of thought, one might similarly attribute the appearance of downward curvature in $S_{ab}(T)$ at the close temperature $T_\mathrm{S}$ to the onset of Fermi-surface reconstruction created by CDW modulations, especially considering that a decrease of $S_{ab}(T)$ and its sign change into negative values have also been observed in the field-induced normal state in underdoped YBCO. \cite{Chang2010,Laliberte2011} Within this scenario, the curvature change of $\rho_{ab}(T)$ at $T_R$ and the subsequent development of $T^2$ dependence below $T_2$ could be viewed as a manifestation of Fermi liquid-like transport behavior of the electronic states of the reconstructed Fermi surface yielded by charge density modulations. \cite{FRemark} For doping decreasing from about 0.10, while a weakening of the CDW order is observed and the $T_{\mathrm{CDW}}$ is shown to decrease \cite{Hucker2014,Blanco-Canosa2014} and become progressively lower than other characteristic temperatures shown in Fig. \ref{fig-various-T-comparison}(b), it has been detected that a spin-density-wave phase gradually emerges and a modulation in spin correlations sets in at the temperature $T_{\mathrm{ELC}}$, which becomes increasingly higher. \cite{Coneri2010,Hinkov2008,Haug2010} Given the good agreement among $T_{\mathrm{ELC}}$, $T_2$, and $T_\mathrm{S}^{\mathrm{max}}$, one may speculate that in this doping region there is a close relationship between the onset of modulation of the spin system and the characteristic changes in the three charge transport properties. As to the temperature $T_{\mathrm{infrared}}$, its agreement with other characteristic temperatures as shown in Fig. \ref{fig-various-T-comparison}(b) over a large range of doping suggests that, in parallel with the curvature change in in-plane transport coefficients, the intra-bilayer anomalies and related anomalies in certain phonon modes observed in the measurement of infrared $c$-axis conductivity \cite{Dubroka2011} may also be linked to the appearance of modulated charge or spin correlations in YBCO, with the phonon anomalies resulted presumably via electron-phonon coupling. \cite{Reznik2006} One may note that this is different from the original proposal made in the infrared-conductivity study, which associates $T_{\mathrm{infrared}}$ with the onset of precursor superconductivity. \cite{Dubroka2011} As discussed above, at the temperature $T_f$ much lower than $T_{\mathrm{infrared}}$, the $\rho_{ab}(T)$, $S_{ab}(T)$, and $R_\mathrm{H}(T)$ of underdoped YBCO show simultaneously upturns in their temperature derivatives, which are just attributed to the emergence of superconducting fluctuations. This, and at the same time the correspondence between $T_{\mathrm{infrared}}$ and the temperatures like $T_R$, $T_\mathrm{H}$, and $T_\mathrm{S}$, lead us to make the above suggestion that the $T_{\mathrm{infrared}}$ might be connected with a physical process other than the onset of superconducting fluctuations. Concerning the behavior of $T_{\mathrm{infrared}}$ in deeply underdoped regime, it is interesting to note that the doping level $\sim0.08-0.06$, at which the $T_{\mathrm{infrared}}$ starts to show a decline as $p$ decreases, coincides approximately with the point at which a low-$T$ upturn in $\rho_{ab}$ begins to appear. What could hide behind this coincidence might deserve investigations. To bring the discussions in this paragraph to a close, we would like to stress that overall the above conjectures are put forward primarily based on the phenomenon of the overlap of various characteristic temperatures as shown in Fig. \ref{fig-various-T-comparison}(b), and further studies are certainly required to offer more detailed examinations on such issues.

\section{\label{sec:summary}Summary}

In summary, the in-plane resistivity $\rho_{ab}$ of YBCO has been reviewed by means of plotting the temperature derivative d$\rho_{ab}$/d$T$. In the underdoped regime, the onset of a well-defined upturn in d$\rho_{ab}$/d$T$ at $T>T_c$, which implies a start of bending down of the $\rho_{ab}$ prior to superconducting transition, has been attributed to the emergence of superconducting fluctuations in the normal state. This attribution has been backed by the finding that the onset of the upturn in d$\rho_{ab}$/d$T$ also coincides with the onset of an upturn in d$S_{ab}$/d$T$, which marks the beginning of a rapid decrease of the in-plane thermopower $S_{ab}$ in normal sate, and with the onset of a deviation of the transverse magnetoresistance from its typical normal-state behavior. It is shown that the determined onset temperature for superconducting fluctuations $T_f$ is well below $T^\ast$ and follows a doping dependence similar to $T_c$, delineating a relatively small fluctuation regime about $5-30$ K wide in temperature. A good agreement of the determined $T_f$ with that obtained from a variety of other experimental techniques \cite{Xu2005,Wang2006,RullierAlbenque2006,Bergeal2008,Chang2010,Grbic2011} is also shown, which suggests further a restricted temperature range of superconducting fluctuations in the pseudogap phase of YBCO.

Above $T_f$, the temperature and doping evolution of d$\rho_{ab}$/d$T$ shows that the $\rho_{ab}(T)$ for $p$ less than about 0.13 develops a prominent $T^2$ dependence below the temperature $T_2$, which is about half the $T^\ast$ and decreases roughly in parallel with $T^\ast$ as doping increases. Moreover, it indicates two critical doping points: $p_1\simeq0.08$ below which an insulating-like component characterized approximately by a $\ln(1/T)$ dependence is also involved in $\rho_{ab}(T)$, and $p_2\simeq0.10$ above which in the majority of samples a $T$-linear component is shown to coexist with the $T$-quadratic in $\rho_{ab}(T)$. It is noted that the presence of this additional $T$-linear term in $\rho_{ab}(T)$ has implications on the in-plane magnetotransport of YBCO over the corresponding doping and temperature regime, as discussed in Sec. \ref{subsec:T-square-Res}. Future studies may be directed to a systematic exploration of $\rho_{ab}$ at temperatures below $T_f$, i.e., in the exposed normal state with the suppression of superconductivity by high magnetic fields. This would help, on one hand, to determine how the $T_2$ varies for $p\gtrsim0.13$ and in particular whether it declines to zero at $p\sim0.16$ as one may suspect from an extrapolation of its doping dependence shown in Fig. \ref{fig-Phasediagram}, and on the other hand, to unravel the fate of the $T$-linear term at low temperatures for samples of $p>0.10$ showing its presence at $T_f<T<T_2$, as there are reports indicating that it may not extend down to $T=0$ K.

In examining the in-plane Hall coefficient $R_\mathrm{H}(T)$ and thermopower $S_{ab}(T)$ of YBCO as typically reported by also plotting the temperature derivatives, it is revealed that, for $p\lesssim0.12$, the temperature $T_2$ corresponds with the temperature $T_\mathrm{S}^{\mathrm{max}}$ at which the $S_{ab}(T)$ shows a maximum, and at a higher temperature $T_R$ for $\rho_{ab}(T)$ to exhibit a curvature change (inflection point), the $R_\mathrm{H}(T)$ and $S_{ab}(T)$ also exhibit curvature change, demonstrating a correlation of the inflection point or curvature change in the temperature dependence of the three in-plane transport quantities. It is further noted that, as illustrated in Fig. \ref{fig-various-T-comparison}(b), these characteristic temperatures identified from the three transport properties show closeness over certain doping regions to the temperature $T_{\mathrm{CDW}}$ or $T_{\mathrm{ELC}}$, namely the onset temperature for the CDW order or an electronic-liquid-crystal state as identified respectively in recent x-ray \cite{Hucker2014,Blanco-Canosa2014} or neutron \cite{Haug2010} scattering experiments, suggesting that the concurrent appearance of inflection point or curvature change in the three in-plane transport properties, and in particular the emergence of $T^2$ dependence in $\rho_{ab}$, might be related to the onset of modulations in charge or spin correlations in the pseudogap phase of YBCO, which deserves further investigations.



\begin{acknowledgments}
The experimental help from Xiao-Yu Du, Chen Zhang, and Yan Zhang is gratefully acknowledged. This work has been supported in part by the MOST of China under 973 Program 2011CBA00106, and by the SRF for ROCS, SEM.
\end{acknowledgments}


\bibliography{}

\begin{thebibliography}{87}
\bibitem [{FRe()}]{FRemark}%
  \BibitemOpen
  \href@noop {} {}\bibinfo {note} {We note that, for Hg1201 ($p\simeq0.09$),
  the onset of charge order has also recently been found in Ref.
  \onlinecite{Tabis2014} to coincide with the appearance of $T^2$ dependence in
  $\rho_{ab}(T)$.}\BibitemShut {Stop}%
\end{thebibliography}

\end{document}